\documentclass{JFM-FLM_Au}

\usepackage{amssymb}
\usepackage{xcolor}
\usepackage{graphicx}
\graphicspath{{./figures/}}
\usepackage{dcolumn}
\usepackage{bm}
\usepackage[normalem]{ulem}
\usepackage{cancel}
\usepackage{booktabs}
\usepackage{multirow} 
\usepackage{hyperref}
\usepackage{tikz,xcolor,hyperref}
\usetikzlibrary{calc,fit}

\hypersetup{
    colorlinks=true,
    linkcolor=red,
    citecolor=black,
    urlcolor=blue
}

\newcommand{\bk}{\boldsymbol{\kappa}}

\definecolor{lime}{HTML}{A6CE39}
\DeclareRobustCommand{\orcidicon}{
	\begin{tikzpicture}
	\draw[lime, fill=lime] (0,0)
 		circle [radius=0.16]
 		node[white] {{\fontfamily{qag}\selectfont \tiny ID}};
 	\draw[white, fill=white] (-0.0625,0.095)
 		circle [radius=0.007];
 	\end{tikzpicture}
 	\hspace{-2mm}
}
	
\foreach \x in {A, ..., Z}{\expandafter\xdef\csname orcid\x\endcsname{\noexpand\href{https://orcid.org/\csname orcidauthor\x\endcsname}
			{\noexpand\orcidicon}}
}

\lefttitle{G. Foggi Rota, A. Mazzino and M. E. Rosti}
\righttitle{Journal of Fluid Mechanics}

\title{Free surfaces in turbulence - A unified framework from water surfaces to elastic solids}

\author{Giulio Foggi Rota \aff{1}\orcidA{}, Andrea Mazzino \aff{2,3}\orcidB{} \and Marco Edoardo Rosti \aff{1}\orcidC{}}

\affiliation{\aff{1}Complex Fluids and Flows Unit, Okinawa Institute of Science and Technology Graduate University (OIST), 1919-1 Tancha, Onna, Okinawa, 904-0495, JP.
\aff{2}Department of Civil, Chemical and Environmental Engineering (DICCA), University of Genova, Via Montallegro 1, Genova, 16145, IT.
\aff{3} National Institute for Nuclear Physics (INFN), Section of Genova, Via Dodecaneso 33, Genova, 16146, IT.}

\corresau{A. Mazzino: andrea.mazzino@unige.it; M. E. Rosti: marco.rosti@oist.jp} 

\begin{document}
\maketitle

\begin{abstract}
What do the ocean surface and a swaying flag have in common? Both are deformable surfaces exhibiting chaotic motion when exposed to turbulent flows. Whether such motion is primarily driven by flow turbulence or by nonlinear dynamics intrinsic to the surface remains debated. Surface waves can interact nonlinearly and transfer energy across scales through the cascade of \textit{wave turbulence}, a behaviour observed at interfaces between otherwise quiescent fluids and in controlled laboratory experiments. They can as well induce turbulent motions in the neighbouring fluids (\textit{wave-induced-turbulence}), provided the local Reynolds number is large enough. Realistic environments, however, are more complex and typically involve the simultaneous presence of wave turbulence and wave-induced-turbulence with \textit{turbulence-induced-waves}, the dynamic relevance of which remains unclear. Here we develop a theoretical framework describing the response of a deformable surface to pressure fluctuations generated by a turbulent flow, and validate it using numerical simulations of the air--water interface in quasi-realistic conditions, complemented by simulations of a deformable rubber layer. Our linear theory, which excludes nonlinear wave--wave interactions, predicts distinct dynamical regimes depending on whether intrinsic surface dynamics emerge or whether the interface is enslaved by flow turbulence. 
Remarkably, although our fully resolved and nonlinear simulations do not inhibit the onset of wave turbulence, \textcolor{black}{interface spectra measured in the configurations studied here are controlled at leading order by turbulent pressure forcing and linear surface response, in strong agreement with our theoretical assumptions. Our predictions appear compatible with aerial surveys of the ocean surface}, highlighting the need for further,  \textcolor{black}{joint measurements of the pressure, elevation, and velocity at the surface} to distinguish among wave turbulence and turbulence-induced-waves.
\end{abstract}


\section{Introduction}
Water is at the origin of life and a fundamental element for human survival and development. Only after learning to navigate the waves of the ocean were humans able to exploit its potential and explore the world. Close to the water surface, buzzing with a multitude of biological processes \citep{sieburth-etal-1976, carlson-mayer-1980}, the complex interplay between surface winds and oceanic currents takes place on a variety of scales. Locally, fluid mixing \citep{zappa-etal-2007} and wave breaking \citep{peregrine-1983, perlin-choi-tian-2013, romero-2019, deike-2022} enhance gas exchange with the atmosphere \citep{liss-slater-1974} (e.g, carbon dioxide absorption in the water). On geophysical scales, evaporation and momentum exchange between the atmosphere and the ocean drive intense climatic events \citep{li-zahariev-garrett-1955,hu-etal-2015,caesar-etal-2018}. Severe weather phenomena such as typhoons and hurricanes pose a significant hazard to human survival \citep{li-chakraborty-2020}, while tsunamis are becoming increasingly dangerous with growing sea levels \citep{chua-etal-2024, shi-etal-2024}. From an engineering perspective, winds, waves, and currents offer potential avenues for energy production \citep{egbert-ray-2000, gonzalez-etal-2024, lu-etal-2024, robertson-etal-2025}.

Air--water interactions in the upper ocean layer, and the dynamics of the interface between them, often occur at high Reynolds numbers, where inertial forces dominate over viscous effects, and the resulting fluid motion is chaotic and turbulent \citep{hartline-1979, smith-thorpe-graham-1999, jones-toba-2001, dasaro-etal-2011}. Understanding how flow turbulence and surface waves interact is therefore of central importance. Although subsurface turbulence often triggers complex surface dynamics \citep{savelsberg-vandewater-2008, savelsberg-vandewater-2009}, the precise nature of these motions remains unclear. In particular, it is still an open question when the surface deforms primarily under the direct influence of flow turbulence, described by Kolmogorov's K41 theory \citep{kolmogorov-1941}, and when it develops a distinct form of turbulent motion---the so-called \textit{wave turbulence} \citep{newell-rumpf-2011, falcon-mordant-2022}---arising from nonlinear wave--wave interactions.

Several kinds of waves have been observed on the ocean surface and multiple theoretical approaches have been proposed to describe their chaotic dynamics \citep{craik-2004, nazarenko-lukaschuk-2016}, ranging from weakly nonlinear wave turbulence theories, where resonant wave--wave interactions drive energy cascades across scales \citep{zakharov-filonenko-1967, newell-rumpf-2011, nazarenko-lukaschuk-2016, falcon-mordant-2022}, to models in which the interface is forced directly by turbulent pressure fluctuations in the flow \citep{bhunia-lienhard-1994}, leading to Phillips and Toba-type equilibrium ranges in the wave spectrum \citep{phillips-1957, toba-1973-2, philips-1985, ryabkova-etal-2019}. \textcolor{black}{A linear-response framework was developed by \citet{perrard-etal-2019} to describe the deformation of a liquid surface subject to turbulent forcing in wind-driven conditions.}

\textcolor{black}{\citet{brocchini-peregrine-2001} classified the possible surface responses in terms of Froude and Weber numbers based on the integral scale and RMS velocity of the underlying flow turbulence, emphasising the competition between turbulent forcing and gravitational and capillary restoring effects. More recently, dedicated experimental and numerical studies have sought to relate observable surface features to the underlying turbulent flow \citep{ruth-coletti-2026,babiker-etal-2026,ferran-etal-2026}. These studies  demonstrate that surface deformation contains quantitative information about subsurface turbulence, while also highlighting the importance of the near-surface flow and, in the presence of a mean current, of Doppler shifting in the observed wave dispersion relation \citep{ferran-etal-2026}.}

When their amplitude is large enough, waves propagating at the water surface might as well induce turbulent motions in the neighbouring fluid, the so-called wave-induced-turbulence \citep{babanin-2006}. However, the relative importance of \textcolor{black}{the mechanisms listed above} remains unsettled. Experiments and simulations are being developed to clarify this, yet each approach has limitations. \textcolor{black}{Laboratory experiments \citep{michel-petrelis-fauve-2017, he-slunyaev-mori-2022, smeltzer-etal-2023, feddersen-etal-2023} have traditionally been restricted by the finite size of the facility and the resulting limited inertial range, although recent investigations have substantially broadened the range of turbulent conditions accessible at the free surface \citep{babiker-etal-2026, ruth-coletti-2026, berhanu-falcon-2026}. Numerical simulations likewise face finite-size effects associated with periodic domains, which limit the range of accessible wavelengths and therefore the scale separation over which asymptotic behaviour can be identified. In addition, many numerical studies \citep{onorato-osborne-serio-2006, baronio-etal-2014, dyachenko-kachulin-zakharov-2017}} often solve Schr\"odinger-like equations \citep{lo-mei-1985,dysthe-1997} that neglect viscous effects and strong nonlinearities, potentially hindering---or even preventing---the emergence of turbulent dynamics. Only in recent years fully resolved, nonlinear multiphase simulations have been performed to study wave breaking \citep{devita-verzicco-iafrati-2018, chan-etal-2021, wu-popinet-deike-2022, digiorgio-pirozzoli-iafrati-2022} and propagation \citep{zonta-soldati-onorato-2015, matsuda-etal-2023, giamagas-etal-2023}, with the latter often yielding results in apparent contradiction with earlier experimental observations \textcolor{black}{due to the inherent difficulty of matching the regimes where simulations and experiments are performed.}
 
Despite this progress, most existing approaches focus either on intrinsic surface dynamics, as in wave turbulence theories, or on their direct forcing by turbulence, and a general framework that consistently couples bulk turbulent motions to the dynamics of a deformable interface (i.e., \textit{turbulence-induced-waves}) is still missing.

\begin{figure}
  \centering
  \includegraphics[width=\textwidth]{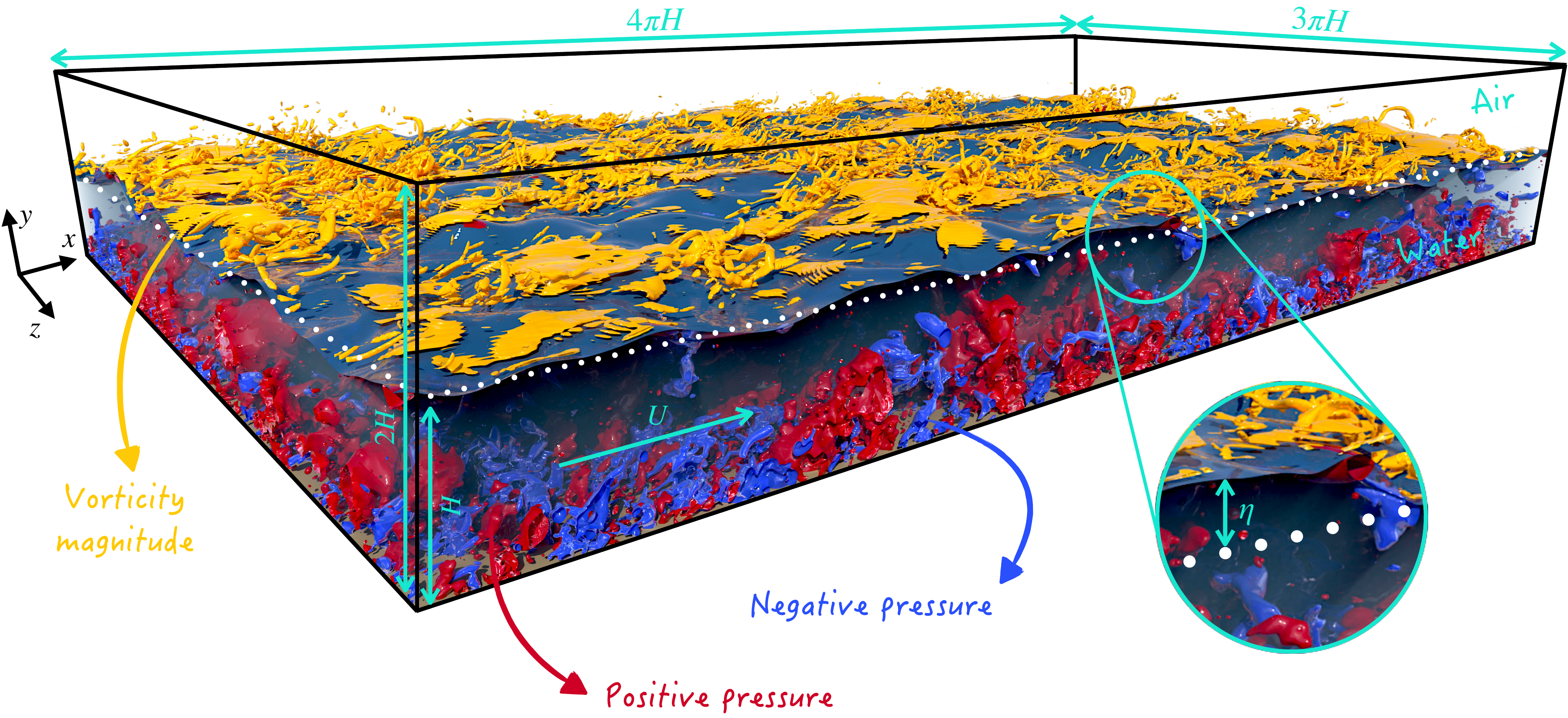}
  \caption{\textbf{Interplay between flow turbulence and surface waves.} We visualise the multiphase {air--water} flow over a smooth bed, as captured by our fully resolved {nonlinear} simulations. Turbulence develops in the water due to friction at the bed and spreads towards the surface, agitated by pressure fluctuations. Surface waves develop and propagate, perturbing the {overlying} wind through the formation of hairpin-like vortical structures. The dynamics of air and water is thus fully turbulent, as well as that of the surface separating them, which is the subject of our investigation.}
  \label{fig:visualisation}
\end{figure}

Here we take a fundamental step forward in understanding the interplay between flow turbulence and surface dynamics by establishing the theoretical foundations of the interaction. 
\textcolor{black}{Our theory applies when the direct response to turbulent pressure fluctuations, together with the linear restoring and damping mechanisms, controls the surface statistics on the timescales considered, while nonlinear spectral redistribution remains sub-leading. It ceases to apply when the nonlinear transfer becomes dynamically comparable to these processes. Under these constraints, we isolate} two distinct dynamical regimes for surface motion. In the first, intrinsic surface dynamics are faster than the surrounding flow turbulence, and emerge. {In the second they are slower, so the surface is \textit{enslaved} by the turbulent forcing and exhibits either an inertia-dominated or a viscous-dominated response depending on the system parameters.}
 {We validate our theoretical predictions by performing state-of-the-art, fully resolved nonlinear simulations.} The quasi-realistic, turbulence-driven wave system shown in figure~\ref{fig:visualisation} conforms to our theoretical predictions in the regime where surface dynamics outpace the turbulent forcing. {When surface dynamics are slower, by contrast, fluid--fluid interfaces tend to break; we therefore support the theory with measurements performed on the surface of a hyperelastic solid exposed to a turbulent flow, used specifically to test the slow, viscous-dominated enslaved limit in which the precise form of the restoring mechanism does not enter at leading order.} We therefore not only validate our theoretical framework introducing state-of-the-art simulations, but also show that its scope extends beyond the water-surface dynamics for which it was conceived: this work provides a powerful, unifying framework for the dynamics of free surfaces in turbulence.

\section{Theoretical framework}

Consider the interface separating two immiscible and incompressible fluids of density $\rho_1,\rho_2$ and kinematic viscosity $\nu_1,\nu_2$, respectively.
\textcolor{black}{Subscript 1 denotes the lighter fluid phase over the interface and subscript 2 the heavier one below.}
Decomposing the interfacial displacement $\eta$ in its Fourier modes $\hat\eta_{\boldsymbol{\kappa}}$ at each wavenumber $\boldsymbol{\kappa}$,
\begin{equation}
{\eta(x,z,t)=\iint \hat\eta_{\boldsymbol{\kappa}}(t)
            e^{\mathrm i(\kappa_xx+\kappa_zz)}\,\mathrm d^2\kappa, \text{ with } \boldsymbol{\kappa}=(\kappa_x,\kappa_z),\; \kappa=|\boldsymbol{\kappa}|,}
\end{equation}
the linearised equation \citep{phillips-1957,li-shen-2023} for each mode writes as a canonical forced and damped oscillator (details on the derivation in appendix~\S\ref{app:osc}),
\begin{equation}
\ddot{\hat\eta}_{\boldsymbol{\kappa}} +2\frac{(\rho_1\nu_1+\rho_2\nu_2)\kappa^2}{\rho_1+\rho_2}\,\dot{\hat\eta}_{\boldsymbol{\kappa}} +\frac{\bigl[(\rho_2-\rho_1)g\kappa+\sigma \kappa^{3}\bigr]}{\rho_1+\rho_2}\hat\eta_{\boldsymbol{\kappa}} =-\frac{\kappa}{\rho_1+\rho_2}\,\widehat{\Delta p'}_{\boldsymbol{\kappa}}(t).
\label{eq:osc}
\end{equation}
Here, dotted quantities have been differentiated in time, $g$ represents the magnitude of the gravity vector, $\sigma$ is the surface tension, and $\widehat{\Delta p'}$ is the fluctuating part of the pressure jump across the interface. We identify the coefficients of equation~\eqref{eq:osc} with twice the damping and the squared natural frequency of the system, respectively
\begin{equation}
\boxed{\gamma(\kappa)={(\rho_1\nu_1+\rho_2\nu_2)\kappa^{2}}/({\rho_1+\rho_2})} \quad \text{and} \quad \boxed{\omega_0^{2}(\kappa)={[(\rho_2-\rho_1)g\,\kappa+\sigma \kappa^{3}]}/({\rho_1+\rho_2}).}
 \label{eq:dispersion}
\end{equation}
\textcolor{black}{Equation~\eqref{eq:osc} can therefore be interpreted as a forced resonant oscillator: for each wavenumber $\kappa$, the natural frequency $\omega_0(\kappa)$ is set by the gravity--capillary dispersion relation, while $\gamma(\kappa)$ controls the dissipative broadening of the resonance. The turbulent pressure fluctuations then provide the broadband forcing that excites this response. Fluctuating tangential stresses at the interface would provide an additional forcing term in the linear response, as considered by \citet{perrard-etal-2019}. Such a contribution is relevant for shear-dominated configurations, e.g., wind-driven waves, whereas the present work focuses on interfaces driven by pressure fluctuations generated by turbulence in the flow. This framework should therefore not be extrapolated to describe shear-driven surface dynamics in its present form, while extensions in the spirit of \cite{perrard-etal-2019} are possible yet out of scope. Additionally, the derivation above is a linearised description and does not contain nonlinear wave--wave interactions, formally assuming  $\kappa\eta<1$ at each scale of interest. It is therefore not intended to describe strongly nonlinear or breaking waves, nor wave turbulence regimes in which resonant interactions provide the leading transfer of energy across scales.}

Applying Fourier's transform in time to equation~\eqref{eq:osc} \textcolor{black}{according to the the convention
\begin{equation}
\widehat{\eta}_{\boldsymbol{\kappa}}(\omega)
=
\int_{-\infty}^{\infty}
\widehat{\eta}_{\boldsymbol{\kappa}}(t)
e^{i\omega t}\,\mathrm{d}t,
\end{equation}
with inverse transform
\begin{equation}
\widehat{\eta}_{\boldsymbol{\kappa}}(t)
=
\frac{1}{2\pi}
\int_{-\infty}^{\infty}
\widehat{\eta}_{\boldsymbol{\kappa}}(\omega)
e^{-i\omega t}\,\mathrm{d}\omega,
\end{equation}
}
we find
\begin{equation}
\hat\eta_{\boldsymbol{\kappa}}(\omega)= \frac{-\kappa\,\widehat{\Delta p'}_{\boldsymbol{\kappa}}(\omega)}
     {(\rho_1+\rho_2)\!\bigl[-(\omega^{2}-\omega_0^{2})-2\mathrm i\gamma\omega\bigr]}
\end{equation}
in the domain of the frequencies $\omega$, and isolate the transfer function 
\begin{equation}
|H(\kappa,\omega)|^{2}\equiv \frac{1}{(\rho_1+\rho_2)^{2}} \frac{\kappa^2\,}{\bigl(\omega^{2}-\omega_0^{2}\bigr)^{2}+4\gamma^{2}\omega^{2}}.
\label{eq:H}
\end{equation}
The azimuthally integrated (isotropic) power spectral density of the {surface} displacement, $S_\eta$, therefore relates to that of the pressure jump, $S_{\Delta p}$, through
\begin{equation}
S_\eta(\kappa,\omega)=|H(\kappa,\omega)|^{2}\,S_{\Delta p}(\kappa,\omega).
 \label{eq:transfer}
\end{equation}
{Here and below, the notation $S(\kappa,\omega)$ denotes a spatiotemporal spectrum, while $S(\kappa)\equiv \int_{-\infty}^{\infty} S(\kappa,\omega)\,d\omega$ denotes the corresponding spectrum integrated over frequency.}
Similarly, the surface vertical velocity, $S_{\dot\eta}$, follows as
\begin{equation}
{S_{\dot\eta}(\kappa,\omega) =\omega^2\,\lvert H(\kappa,\omega)\rvert^2\,S_{\Delta p}(\kappa,\omega).}
\label{eq:dotTransfer}
\end{equation}

\subsection{Kolmogorov's spectrum}

 Kolmogorov's theory \textcolor{black}{\citep{kolmogorov-1941,monin-yaglom-1971}} allows to predict a one-dimensional pressure spectrum of the form
\begin{equation}
{S_{\Delta p}(\kappa) = C_p\,\rho^2\,\epsilon^{4/3}\;\kappa^{-7/3},}
\label{eq:pressureK41}
\end{equation}
where $\rho$ and $\epsilon$ denote the effective density and turbulent dissipation rate of the flow at the surface, respectively
\begin{equation}
\rho^2 \equiv \rho_1^2 + \rho_2^2, \text{ and } \epsilon^{4/3} \equiv ({\rho_1^2\,\epsilon_1^{4/3} + \rho_2^2\,\epsilon_2^{4/3}})/({\rho_1^2 + \rho_2^2}).
\end{equation}
Introducing the eddy turnover time (and its inverse) as
\begin{equation}
\tau_r(\kappa)=\epsilon^{-1/3}\kappa^{-2/3}, \quad \boxed{\omega_t(\kappa)=1/{\tau_r(\kappa)}=\epsilon^{1/3}\kappa^{2/3},}
\end{equation}
the full spatiotemporal spectrum is then written as
\begin{equation}
{S_{\Delta p}(\kappa,\omega) = S_{\Delta p}(\kappa)\;\frac{1}{\omega_t(\kappa)}\;
  F \Bigl(\frac{\omega}{\omega_t(\kappa)}\Bigr), \qquad \int_{-\infty}^{\infty}F(\xi)\,d\xi = 1.}
  \label{eq:closure}
\end{equation}
\textcolor{black}{Equation~\eqref{eq:closure} is a similarity closure for the frequency distribution of the pressure forcing at fixed wavenumber, rather than a one-to-one correspondence $\omega=\omega_t(\kappa)$. At fixed $\kappa$, $F$ is the normalised distribution of pressure-fluctuation variance over frequency and it is assumed to have its dominant weight for $|\omega|=O(\omega_t)$. Taylor advection and random sweeping instead concern the Eulerian frequency broadening observed at a fixed point. They are not used here to convert a spatial spectrum into a temporal one: the full $\omega$ dependence is retained and the response spectra are subsequently integrated over frequency. Mean Doppler shifting associated with interface advection is accounted for in the space--time analysis of figure~\ref{fig:results}(c).} {For the asymptotic estimates below we assume that \textcolor{black}{$F$ decays ``sufficiently'' fast and has negligible weight in the $|\omega-\omega_0|\lesssim\gamma$ band, and that its} low-frequency core is sufficiently regularised, for instance $F(\xi)=O(|\xi|^{\alpha})$ as $\xi\to0$ with $\alpha>3$. Under these assumptions, both $\int F(\xi)\,\xi^{-4}\,d\xi$ and $\int F(\xi)\,\xi^{-2}\,d\xi$ are finite constants; only these pre-factors depend on the detailed shape of $F$, whereas the $\kappa$-scalings derived below do not.}

{
\subsection{Regimes of motion of the surface - Pressure balance}

A physically transparent classification of the regimes of motion of the surface follows directly from equation~\eqref{eq:osc}. Multiplying it by $(\rho_1+\rho_2)/\kappa$, each Fourier mode can be written as a balance between three effective pressure contributions and the turbulent forcing,
\begin{equation}
\underbrace{\frac{\rho_1+\rho_2}{\kappa}\,\ddot{\hat\eta}_{\boldsymbol{\kappa}}}_{\text{inertial pressure}}
+\underbrace{2(\rho_1\nu_1+\rho_2\nu_2)\kappa\,\dot{\hat\eta}_{\boldsymbol{\kappa}}}_{\text{viscous pressure}}
+\underbrace{\bigl[(\rho_2-\rho_1)g+\sigma \kappa^{2}\bigr]\hat\eta_{\boldsymbol{\kappa}}}_{\text{restoring pressure}}
=-\widehat{\Delta p'}_{\boldsymbol{\kappa}}(t).
\label{eq:pressure-balance}
\end{equation}
\textcolor{black}{Consistently with the closure~\eqref{eq:closure}, the estimates below concern the turbulence-forced stationary component whose frequency integrals are controlled by the band $|\omega|=O(\omega_t)$. Accordingly, at the level of scaling,
\begin{equation}
\dot{\hat\eta}_{\boldsymbol{\kappa}} \sim \omega_t(\kappa)\,\hat\eta_{\boldsymbol{\kappa}},
\qquad
\ddot{\hat\eta}_{\boldsymbol{\kappa}} \sim \omega_t^2(\kappa)\,\hat\eta_{\boldsymbol{\kappa}}.
\label{eq:turbulent-modulation}
\end{equation}
These do not identify $\omega_t$ with the natural surface frequency $\omega_0$: the latter remains fully retained in the transfer function~\eqref{eq:H}, while the ratio $\delta=\omega_0/\omega_t$ introduced below quantifies the competition between turbulent forcing and intrinsic surface dynamics.}

The three effective pressures then scale as
\begin{equation}
p_I \sim \frac{\rho_1+\rho_2}{\kappa}\,\omega_t^2\,\hat\eta_{\boldsymbol{\kappa}},
\qquad
p_V \sim (\rho_1\nu_1+\rho_2\nu_2)\kappa\,\omega_t\,\hat\eta_{\boldsymbol{\kappa}},
\qquad
p_R \sim \bigl[(\rho_2-\rho_1)g+\sigma\kappa^2\bigr]\hat\eta_{\boldsymbol{\kappa}}.
\end{equation}
Using equations~\eqref{eq:dispersion} and the definition of $\omega_t$, the ratios between these contributions are immediately identified as
\begin{equation}
\boxed{\delta=\frac{\omega_0(\kappa)}{\omega_t(\kappa)}},
\qquad
\boxed{\zeta=\frac{\gamma(\kappa)}{\omega_t(\kappa)}},
\end{equation}
so that $p_R/p_I \sim \delta^2$ and $p_V/p_I \sim \zeta$.
The interface regimes can therefore be obtained from dominant balances in equation~\eqref{eq:pressure-balance}.
\textcolor{black}{Before presenting them, we note that both $\delta$ and $\zeta$ are inherently scale-dependent, since the intrinsic frequency $\omega_0$, the damping rate $\gamma$, and the turbulent turnover frequency $\omega_t$ all depend on the wavenumber $\kappa$. Thus, any distinction between wave-dominated and turbulence-dominated dynamics is to be understood locally in wavenumber space. In particular, $\delta\gg1$ indicates that intrinsic surface dynamics are faster than the turbulent dynamics and can emerge as surface waves, whereas $\delta\ll1$ corresponds to an interface enslaved by the turbulent flow. Similarly, $\zeta$ compares the viscous relaxation rate of the interface with the timescale of the turbulent forcing and distinguishes the inertia-dominated and viscosity-dominated limits of the turbulence-enslaved regime.}

\begin{itemize}
	\item ${\delta\gg1}$ (intrinsic surface dynamics are faster than turbulence dynamics, and emerge)\\
	In this regime the restoring pressure dominates over inertia on the turbulent timescale, and viscous dissipation remains subleading provided $\zeta \ll \delta^2$, i.e.\ $p_R \gg p_I$ and $p_R \gg p_V$. The leading-order balance is therefore
	\begin{equation}
	\bigl[(\rho_2-\rho_1)g+\sigma\kappa^2\bigr]\hat\eta_{\boldsymbol{\kappa}}
	\sim
	-\widehat{\Delta p'}_{\boldsymbol{\kappa}}.
	\label{eq:restoring-balance}
	\end{equation}
	The surface elevation is thus set by the restoring mechanism active at that scale:
	\begin{equation}
	\hat\eta_{\boldsymbol{\kappa}}
	\sim
	\frac{\widehat{\Delta p'}_{\boldsymbol{\kappa}}}{(\rho_2-\rho_1)g+\sigma\kappa^2},
	\end{equation}
	\textcolor{black}{and the crossover between gravity and capillary waves, regardless of flow turbulence, occurs when \( (\rho_2-\rho_1)\,g\,\kappa = \sigma\,\kappa^3,\) yielding the critical wavenumber 
        \begin{equation}
        \boxed{\kappa_c = \sqrt{{(\rho_2-\rho_1)\,g}/{\sigma}}.}
        \label{eq:transition}
        \end{equation}
        }
        In the gravitational limit, $\kappa\ll\kappa_c$, the restoring pressure is hydrostatic and
	\begin{equation}
	\hat\eta_{\boldsymbol{\kappa}} \sim \frac{\widehat{\Delta p'}_{\boldsymbol{\kappa}}}{(\rho_2-\rho_1)g}
	\quad\Rightarrow\quad
	\boxed{S_{\eta}(\kappa)\propto S_{\Delta p}(\kappa)\propto \kappa^{-7/3}.}
	\end{equation}
	In the capillary limit, $\kappa\gg\kappa_c$, curvature provides the restoring pressure and
	\begin{equation}
	\hat\eta_{\boldsymbol{\kappa}} \sim \frac{\widehat{\Delta p'}_{\boldsymbol{\kappa}}}{\sigma\kappa^2}
	\quad\Rightarrow\quad
	\boxed{S_{\eta}(\kappa)\propto \kappa^{-4}S_{\Delta p}(\kappa)\propto \kappa^{-19/3}.}
	\end{equation}
	Although the amplitude is selected by the intrinsic restoring dynamics, the temporal modulation remains imposed by the turbulent forcing, hence
	\begin{equation}
	S_{\dot\eta}(\kappa)\sim \omega_t^2(\kappa)\,S_{\eta}(\kappa)
	\quad\Rightarrow\quad
	\boxed{
	S_{\dot\eta}(\kappa)\propto
	\begin{cases}
	\kappa^{-1}, & \text{gravitational limit},\\[4pt]
	\kappa^{-5}, & \text{capillary limit}.
	\end{cases}}
	\end{equation}

	\item ${\delta\ll1}$ (surface dynamics are slower than turbulence dynamics, and the surface is \textit{enslaved} by flow turbulence)\\
	The restoring pressure is now too slow to react on the turbulent timescale, so the response is controlled by whichever term, inertia or viscous dissipation, balances the pressure forcing most efficiently.
	\begin{itemize}
  		\item{\textit{$\zeta\ll1$ (inertia-dominated enslaved limit)}}: viscous effects are weak on the turbulent timescale, so the forcing balances the effective interfacial inertia,
		\begin{equation}
		\frac{\rho_1+\rho_2}{\kappa}\,\ddot{\hat\eta}_{\boldsymbol{\kappa}}
		\sim
		-\widehat{\Delta p'}_{\boldsymbol{\kappa}}.
		\end{equation}
		Using equation~\eqref{eq:turbulent-modulation},
		\begin{equation}
		\hat\eta_{\boldsymbol{\kappa}}
		\sim
		\frac{\kappa\,\widehat{\Delta p'}_{\boldsymbol{\kappa}}}{(\rho_1+\rho_2)\omega_t^2}
		\quad\Rightarrow\quad
		\boxed{
		S_{\eta}(\kappa)
		\propto
		\frac{\kappa^2}{\omega_t^4}\,S_{\Delta p}(\kappa)
		\propto
		\kappa^{-3}.}
		\end{equation}
		Since the dominant temporal scale is still $\omega_t$,
		\begin{equation}
		\boxed{
		S_{\dot\eta}(\kappa)
		\propto
		\omega_t^2\,S_{\eta}(\kappa)
		\propto
		\kappa^{-5/3}.}
		\end{equation}

  		\item{\textit{$\zeta\gg1$ (viscous-dominated enslaved limit)}}: viscous dissipation reacts faster than inertia, so the forcing balances the viscous pressure,
		\begin{equation}
		2(\rho_1\nu_1+\rho_2\nu_2)\kappa\,\dot{\hat\eta}_{\boldsymbol{\kappa}}
		\sim
		-\widehat{\Delta p'}_{\boldsymbol{\kappa}}.
		\end{equation}
		Therefore
		\begin{equation}
		\dot{\hat\eta}_{\boldsymbol{\kappa}}
		\sim
		\frac{\widehat{\Delta p'}_{\boldsymbol{\kappa}}}{(\rho_1\nu_1+\rho_2\nu_2)\kappa}
		\quad\Rightarrow\quad
		\boxed{
		S_{\dot\eta}(\kappa)
		\propto
		\kappa^{-2}S_{\Delta p}(\kappa)
		\propto
		\kappa^{-13/3}.}
		\end{equation}
		Since $\hat\eta_{\boldsymbol{\kappa}}\sim \dot{\hat\eta}_{\boldsymbol{\kappa}}/\omega_t$ on the forcing timescale,
		\begin{equation}
		\boxed{
		S_{\eta}(\kappa)
		\propto
		\omega_t^{-2}\,S_{\dot\eta}(\kappa)
		\propto
		\kappa^{-17/3}.}
		\end{equation}
	\end{itemize}
\end{itemize}

\textcolor{black}{We finally note that the parameter $\delta$ connects naturally with the phenomenological classification proposed by \citet{brocchini-peregrine-2001}. Indeed, in the gravity-dominated regime, evaluating $\delta$ at the scale $\kappa\sim H^{-1}$ and estimating the dissipation as $\epsilon\sim u_{\rm rms}^3/H$, where $u_{\rm rms}$ is the RMS turbulent velocity, gives
\begin{equation}
\delta \sim
\frac{1}{u_{\rm rms}}\sqrt{(\rho_2-\rho_1)gH/(\rho_2+\rho_1)}
\sim Fr_H^{-1},
\end{equation}
where $Fr_H=u_{\rm rms}/\sqrt{gH}$ is the integral-scale Froude number of \citet{brocchini-peregrine-2001}. Thus, the present formulation recovers their gravity-dominated similitude when evaluated at the integral scale, while allowing the relative importance of turbulent and intrinsic surface dynamics to be assessed at arbitrary wavenumbers.}

\subsection{Regimes of motion of the surface - Formal derivation}

We now demonstrate that the scaling laws attained above from pressure balance arguments can be formally derived from the integral version of equations \eqref{eq:transfer} and \eqref{eq:dotTransfer}, e.g.,
\begin{equation}
S_{\eta}(\kappa) = \int_{-\infty}^\infty |H(\kappa,\omega)|^2 \;S_{\Delta p}(\kappa,\omega)\,d\omega.
\end{equation}
Consider the transformation
\begin{equation}
\omega=\omega_t(\kappa)\,\xi,
\label{eq:rescaling}
\end{equation}
which yields
\begin{equation}
S_{\eta}(\kappa) \;\propto\; \kappa^{-7/3}\,\frac{\kappa^2}{\omega_t^4} \int_{-\infty}^\infty \frac{F(\xi)\,d\xi}{(\xi^2-\delta^2)^2 + 4\,\zeta^2\,\xi^2} \;=\; \kappa^{-3}\,I(\delta,\zeta),
\end{equation}
where
\begin{equation}
I(\delta,\zeta)=\int_{-\infty}^\infty \frac{F(\xi)\,d\xi}{(\xi^2-\delta^2)^2 + 4\,\zeta^2\,\xi^2}.
\label{eq:spectrum}
\end{equation}
\textcolor{black}{The rescaling~\eqref{eq:rescaling} does not impose a one-to-one relation between frequency and wavenumber; it only measures frequency in units of the characteristic (turbulent) forcing frequency. The surface dispersion relation enters independently through $\omega_0(\kappa)$, or equivalently through $\delta=\omega_0/\omega_t$.}

We now recover the same asymptotic regimes through the integral $I(\delta,\zeta)$, under the regularity assumption on $F$ stated above:
\begin{itemize}
	\item ${\delta\gg1}$ (intrinsic surface dynamics are faster than turbulence dynamics, and emerge)\\
	 	 In this regime, together with $\zeta \ll \delta^2$, one has $(\xi^2-\delta^2)^2 \;\simeq\;\delta^4$ and $4\zeta^2\xi^2 \ll \delta^4$ over the forcing range, so $I\approx {1}/{\delta^4} = {\omega_t^4}/{\omega_0^4}$, hence
	\begin{equation}
	\boxed{S_{\eta}(\kappa)
	\propto
	\kappa^{-3}\,\frac{\omega_t^4}{\omega_0^4}
	\propto
	\begin{cases}
	\kappa^{-7/3} & \text{ when } \omega_0^2\propto g\,\kappa \text{ (gravitational limit)},\\[8pt]
	\kappa^{-19/3} & \text{ when } \omega_0^2\propto \sigma\,\kappa^3 \text{ (capillary limit)}.
	\end{cases}
	}
    \label{eq:Seta}
	\end{equation}
	The gravitational limit again shows a spectrum coinciding with the 1D pressure spectrum.
	 
	\item ${\delta\ll1}$ (surface dynamics are slower than turbulence dynamics, and the surface is \textit{enslaved} by flow turbulence)\\
	 	We further distinguish two sub-limits, consistent with the physically driven balances above.
	\begin{itemize}
  		\item{\textit{$\zeta\ll1$ (inertia-dominated enslaved limit)}}: $(\xi^2 - \delta^2)^2 \approx \xi^4, \quad 4\,\zeta^2 \xi^2 \ll \xi^4, \quad I \approx {\displaystyle \int \frac{F(\xi)}{\xi^4}\,d\xi} = \text{const}$, hence
		\begin{equation}
		\boxed{
		S_{\eta}(\kappa)
		\;\propto\;
		\kappa^{-3}.}
		\end{equation}
     		
  		\item{\textit{$\zeta\gg1$ (viscous-dominated enslaved limit)}}: $(\xi^2-\delta^2)^2 \ll 4\,\zeta^2\xi^2, \quad I \approx {\displaystyle \int \frac{F(\xi)}{4\,\zeta^2\,\xi^2}\,d\xi} = \displaystyle\frac{1}{4\,\zeta^2}\,{\displaystyle \int \frac{F(\xi)}{\xi^2}\,d\xi} \;\propto\;\frac{1}{\zeta^2},$ and since \(\zeta\propto \kappa^{4/3}\), we obtain
		\begin{equation}
		\boxed{
		S_{\eta}(\kappa)
		\;\propto\;
		\kappa^{-3}\,\frac{1}{\zeta^2}
		\;\propto\;
		\kappa^{-17/3}.}
		\end{equation}
	\end{itemize}
\end{itemize}
Since the forcing remains concentrated around frequencies $\omega=O(\omega_t)$, the scaling behaviors of $S_{\dot{\eta}}(\kappa)$ follow consistently from the relationship $ S_{\dot\eta}(\kappa) \sim \omega_t^2\,S_\eta(\kappa)$ and are summarised in table~\ref{tab:summary-scaling} along with those for $S_{\eta}(\kappa)$.
}

\begin{table}
\centering
{
\renewcommand{\arraystretch}{1.3}
\begin{tabular}{@{}l|cccc@{}}
\toprule
\textbf{Regime} & \textbf{Dominant balance} & \textbf{Spectrum} & \textbf{Gravity} & \textbf{Capillary} \\
\midrule
\multirow{2}{*}{\shortstack[l]{$\delta \gg 1,\ \color{black}{\zeta \ll \delta^2}$ \\ emergence of \textit{intrinsic surface dynamics}}} & \multirow{2}{*}{restoring force} & Elevation \(S_\eta(\kappa)\) & \(\kappa^{-7/3}\) & \(\kappa^{-19/3}\) \\
&& Velocity \(S_{\dot\eta}(\kappa)\) & \(\kappa^{-1}\) & \(\kappa^{-5}\) \\
\addlinespace \hline
\vspace{-0.3cm} \\
\multirow{2}{*}{\shortstack[l]{$\delta \ll 1,\ \zeta \ll 1$ \\ inertia-dominated \\ \textit{enslaved} limit}} & \multirow{2}{*}{inertia} & Elevation \(S_\eta(\kappa)\) & \multicolumn{2}{c}{\(\kappa^{-3}\)} \\
&& Velocity \(S_{\dot\eta}(\kappa)\) & \multicolumn{2}{c}{\(\kappa^{-5/3}\)} \\
\addlinespace \hline
\vspace{-0.3cm} \\
\multirow{2}{*}{\shortstack[l]{$\delta \ll 1,\ \zeta \gg 1$ \\ viscous-dominated \\ \textit{enslaved} limit}} & \multirow{2}{*}{viscosity} & Elevation \(S_\eta(\kappa)\) & \multicolumn{2}{c}{\(\kappa^{-17/3}\)} \\
&& Velocity \(S_{\dot\eta}(\kappa)\) & \multicolumn{2}{c}{\(\kappa^{-13/3}\)} \\
\bottomrule
\end{tabular}
\caption{\textbf{Summary of the asymptotic scaling regimes for the 1D surface elevation spectrum \(S_\eta(\kappa)\) and the 1D surface velocity spectrum \(S_{\dot\eta}(\kappa)\) in response to Kolmogorov's pressure forcing.} When intrinsic surface dynamics emerge, the leading-order balance is provided by the restoring force, and the resulting scaling depends on whether gravity or capillarity supplies that restoring action. When the surface is enslaved by turbulence, by contrast, the response becomes either inertia-dominated (\(\zeta \ll 1\)) or viscous-dominated (\(\zeta \gg 1\)), independently of the restoring-force type.}
\label{tab:summary-scaling}
}
\end{table}

\section{Numerical validation}

To validate the theoretical framework we developed to relate flow turbulence and surface motion, we now proceed with fully resolved, nonlinear, and coupled multiphase simulations enforcing the governing balances of incompressible fluid motion, i.e.\ the Navier--Stokes equations:
\begin{equation}
   \displaystyle \frac{\partial \mathbf{u}}{\partial t} + \nabla \cdot (\mathbf{u}\mathbf{u})
   = \frac{1}{\rho_f}\left(- \nabla p + \nabla \cdot \mathbf{\mathcal{T}}\right) + \mathbf{f} + \mathbf{g},
    \qquad \nabla \cdot \mathbf{u} = 0.
   \label{eq:NS}
\end{equation}
Here, $\mathbf{u}$ and $p$ denote the velocity and pressure fields, $\rho_f$ is the local fluid density, $\mathbf{\mathcal{T}}$ is the stress tensor, $\mathbf{f}$ accounts for prescribed volume forces, and $\mathbf{g} = -g\hat{\mathbf{y}}$ is the gravitational acceleration along the negative $\hat{\mathbf{y}}$ direction. Fluid motion is sustained along the streamwise $\hat{\mathbf{x}}$ direction prescribing a forcing $f\hat{\mathbf{x}}$ compatible with the desired turbulent flow regime.

\subsection{Quasi air--water simulations: methods}

Focusing first on the flow regime where intrinsic surface dynamics emerge from the underlying turbulence, we simulate the motion of a stratified {air--water} layer (figure~\ref{fig:visualisation}) under close to realistic conditions.  \textcolor{black}{The flow parameters are chosen to achieve fully developed turbulence in both fluids, ensuring $\delta > 1$ and $\zeta\ll\delta^2$ across all scales of motion (as demonstrated in appendix~\S\ref{app:specs})}.

We consider a computational domain of size $4\pi H \times 2H \times 3\pi H$ (streamwise, vertical, spanwise), as shown in figure~\ref{fig:visualisation}. Its size, informed by previous investigations from the authors \citep{foggirota-chiarini-rosti-2026}, is sufficiently large to allow collection of uncorrelated flow and interface statistics.
\textcolor{black}{All components of the flow velocity are set to zero at the bottom boundary, while at the top boundary we set to zero the vertical velocity and the vertical derivatives of the two remaining velocity components.}
Periodicity is enforced in the streamwise and spanwise directions. The turbulent flow is driven by a constant forcing $f\hat{\mathbf{x}}$, yielding a bulk Reynolds number $Re = \rho_w U H / \mu_w \approx 21{,}000$ based on the water depth $H$, mean velocity $U$, density $\rho_w$, and viscosity $\mu_w$. The bulk Froude and Weber numbers are $Fr = U/\sqrt{gH} = \mathcal{O}(1)$ and $We = \rho_w U^2 H/\sigma = \mathcal{O}(100)$, respectively, where $\sigma$ denotes the surface tension.
\textcolor{black}{For reference, considering standard water properties and an experimental setup with depth of $5$cm, the Reynolds, Froude, and Weber numbers of our simulation would be matched by an average flow speed of $U=0.42\,$m/s.}

The viscosity ratio in the simulations is set to its physical value $\mu_w/\mu_a = 55$, while the density ratio is chosen as $\rho_w/\rho_a = 80$, a compromise enabling numerically stable two-phase simulations. Despite being lower than the physical ratio, the corresponding Atwood number $A = (\rho_w-\rho_a)/(\rho_w+\rho_a) = 0.975$ is only 2\% below the realistic value ($A = 0.997$). Interfacial dynamics saturate for $A \gtrsim 0.94$ \citep{burton-2011}, so no qualitative changes are expected at the physical ratio.

Air and water are distinguished through the \textit{volume-of-fluid} (VOF) colour function $\phi(\mathbf{x},t)$ \citep{rosti-devita-brandt-2019,devita-etal-2021,hori-etal-2023,cannon-soligo-rosti-2024}, and surface tension is incorporated as a body force acting at the interface \citep{popinet-2018} so that $\mathbf{f}=f\hat{\mathbf{x}}+\sigma k \delta_s \hat{\mathbf{n}}$ where $k$ is the local curvature of the interface, $\hat{\mathbf{n}}$ its normal unit vector and $\delta_s$ a Dirac $\delta$-function that is nonzero only on the interface. The problem formulation thus remains monolithic: \textcolor{black}{the governing equations are written in a unified form over the entire computational domain, rather than introducing separate sets of governing equations and separate solution procedures for the individual phases. The presence of multiple phases is accounted for through the phase-dependent material properties, which vary spatially across the interface. Thus, the same conservation equations are solved throughout the domain, without explicitly decomposing the problem into separate fluid subproblems coupled through interfacial conditions.} Local density and viscosity are obtained by cell-averaged interpolation via $\phi$, and the linear constitutive relation $\mathbf{\mathcal{T}} = \mu_f(\nabla\mathbf{u} + \nabla\mathbf{u}^T)$ is retained, with $\mu_f$ the local dynamic viscosity of the fluid.

Equations~\eqref{eq:NS} are discretised on a staggered Cartesian mesh of $2304 \times 480 \times1728$ points, uniform in the horizontal directions and stretched vertically. \textcolor{black}{The vertical grid is chosen so that its first point off the bottom wall lies at $y^+=1$, where the plus apex denotes a quantity scaled with the mean shear at the bottom wall. The spacing is maintained constant up to $y/H\approx0.3$, where it starts to be linearly relaxed so as to attain cubic cells in the interface region, $0.8<y/H<1.2$, where the spacing is once again kept constant. Above that, the grid is further relaxed linearly so as to eventually double the vertical spacing when the top free-slip boundary is reached.}
To confirm that our simulations are capable of resolving the smallest scales of motion we measure the local average value of the Kolmogorov turbulent length-scale ($l_K=\nu^{3/4}\epsilon^{-1/4}$, with $\nu$ and $\epsilon$ the kinematic viscosity and turbulent dissipation rate, respectively) and verify that it remains comparable to the adopted grid spacing $\Delta$ at each location across the vertical extent of the domain, $l_K \approx \Delta$. Furthermore, to asses grid and domain independence, we compare interface statistics from the main study with those attained (i) halving the resolution leaving the domain size unchanged and (ii) halving the domain size along the homogeneous directions, leaving the resolution unchanged. We thus introduce a \textit{half resolution} simulation performed in a box of size $4\pi H \times 2H \times 3\pi H$ discretised over $1152 \times 240 \times 864$ grid points, and a \textit{half domain} simulation performed in a box of size $2\pi H \times 2H \times 1.5\pi H$ discretised over $1152 \times 480 \times 864$ points. Interface statistics are collected after any transient has elapsed and only exhibit negligible departures from each other across simulations, confirming the accuracy of our results in the whole range of scales presented in this study. 

Simulations are performed with the well-validated finite-difference solver \textit{Fujin} maintained by our Unit at OIST \citep[\url{https://groups.oist.jp/cffu/code},][]{rosti-2026}, using second-order central spatial discretisation and a second-order Adams--Bashforth scheme for time advancement. Computations ran on 9{,}216 Fujitsu A64FX CPUs on the supercomputer Fugaku (\url{https://www.r-ccs.riken.jp/en/fugaku/}) for a total of $\sim 25$ million core-hours, \textcolor{black}{spending about $4$ months running and a similar time queuing. The full execution therefore required $8$ months; the same computation, on a single CPU without queuing time, would have required roughly $2800$ years.} 
 A statistically steady state is reached after a transient of $\sim 200\,H/U$, followed by data collection over $\sim 70\,H/U$ using a time step of $2\times10^{-4}\,H/U$.
 \textcolor{black}{Consequently, our simulations cannot rule out the development of a slower wave-kinetic evolution on arbitrarily longer times. Our claims in the following are restricted to this statistically stationary, continuously forced regime. If resonant wave--wave interactions became dominant over substantially longer times or under different forcing conditions, the system would leave the range of applicability of the theory presented earlier.}

\subsection{Quasi air--water simulations: results}

\begin{figure}
  \centering
  \includegraphics[width=\textwidth]{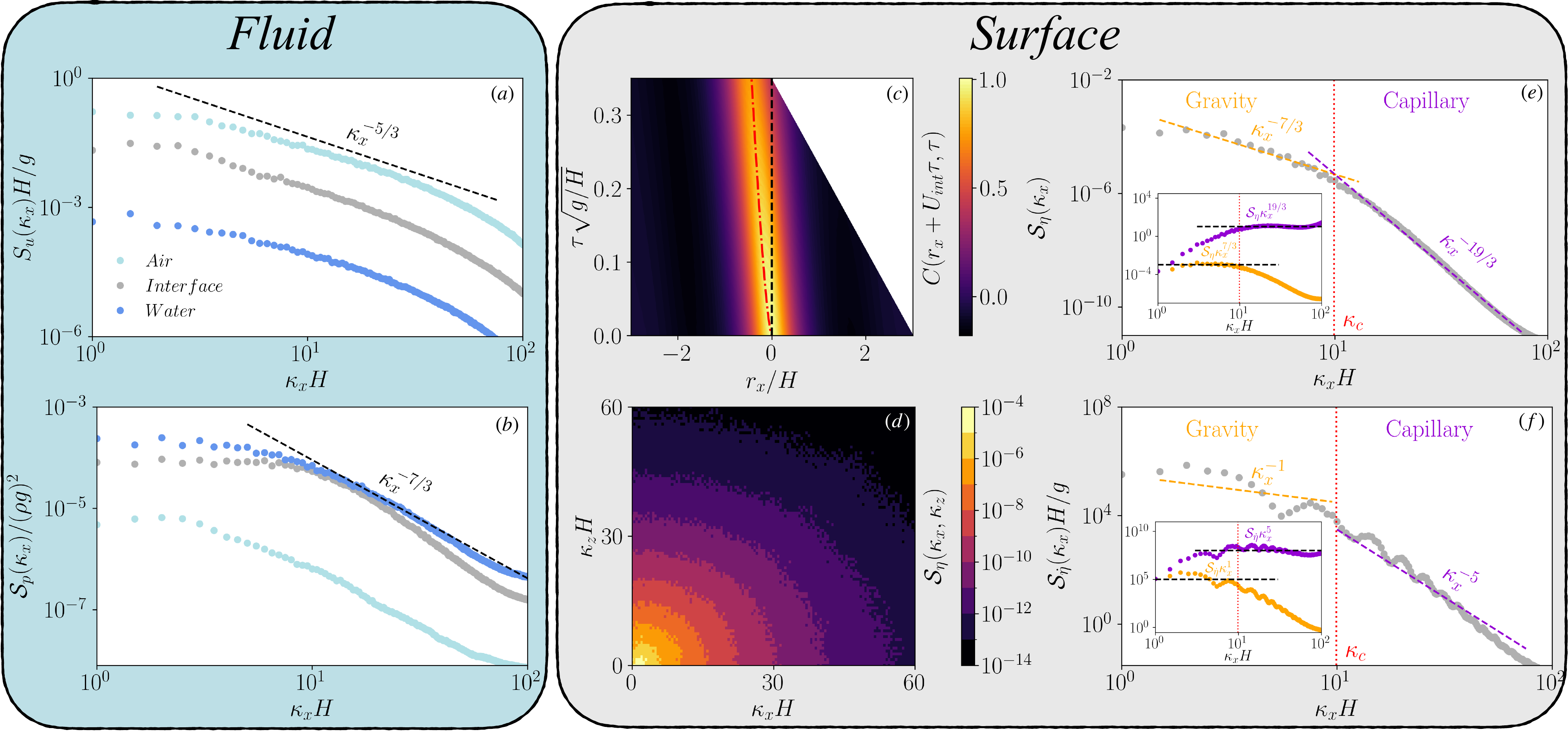}
  \caption{\textbf{Water surface dynamics in turbulence.} Streamwise spectra of the turbulent kinetic energy (panel $a$), arbitrarily shifted to avoid overlapping, and of the pressure (panel $b$), closely matching Kolmogorov's predictions \citep{kolmogorov-1941}. Wave propagation at the surface (panel $c$), observed through the autocorrelation function {$\mathcal{C}(r_x+U_{\rm int}\tau,\tau) = \langle \eta(x,z;t) \eta(x + r_x+U_{\rm int}\tau,z;t+\tau) \rangle/\langle \eta(x,z;t) \eta(x,z;t) \rangle$}, with angle brackets denoting ensemble averaging in time and along the streamwise and spanwise directions, $r_x$ an arbitrary increment in the streamwise direction, $\tau$ an arbitrary increment in time, and $U_{\rm int}$ the mean velocity at which the surface advects downstream. The maxima of $\mathcal{C}$ follow the dispersion relation for {gravity--capillary} waves, reported as a red {dash--dotted} line and made explicit \textcolor{black}{in text}. After confirming that the surface elevation spectrum is almost isotropic (panel $d$), we focus on the streamwise direction and observe the elevation (panel $e$) and vertical velocity spectra (panel $f$) of the water surface (compensated with their predicted scalings in the insets), finding remarkable agreement with our theoretical predictions in the regime where surface dynamics are faster than flow turbulence. \textcolor{black}{This observation not only corroborates the validity of our theory for the considered interface state, but also appears consistent with a sub-dominant role of wave turbulence there, as its signature scalings (\(S_{\eta}\sim \kappa_x^{-15/4}\) and \(S_{\dot\eta}\sim \kappa_x^{-3/4}\) in the capillary regime) are not recovered.}  
  }
  \label{fig:results}
\end{figure}

We verify the turbulent nature of the flow in  appendix~\S\ref{app:specs} and here, by sampling the velocity fluctuations and computing their kinetic energy spectra across streamwise wavenumbers, $S_u(\kappa_x)$, in the water, the air, and at the mean surface position (figure~\ref{fig:results}a). In all cases, we find a well-defined inertial range where $S_u \sim \kappa_x^{-5/3}$\citep{kolmogorov-1941}, extending over \textcolor{black}{about one decade} of wavenumbers. Consistently, the pressure spectrum $S_p$ (figure~\ref{fig:results}b) follows the canonical $S_p \sim \kappa_x^{-7/3}$ scaling on which our theory is built, as per equation~\eqref{eq:pressureK41}. \textcolor{black}{We reported here the pressure spectrum, $S_p$, while our theory relies on the pressure jump across the surface, $S_{\Delta p}$. The two exhibit similar scaling behaviour since pressure fluctuations in the water dominate the jump due to the strong density contrast between air and water, while $S_p$ is more directly accessible. The pressure spectrum is computed as the Fourier transform of the pressure signal along the streamwise direction, averaged in span; yet the scaling behaviour remains unchanged when transforming along the transverse direction instead.}

\textcolor{black}{We then turn to the surface displacement $\eta$, exhibiting undulations of the order $\eta_{\rm rms}\approx 0.07 H$ and ${\partial_{i,j}\eta}_{\rm rms}\approx 0.2$, and examine wave propagation by computing the {space--time} autocorrelation $\mathcal{C}(r_x,\tau)=\langle \eta(x+r_x+U_{\rm int}\tau,z;t+\tau)\,\eta(x,z;t)\rangle/\langle \eta(x,z;t)\,\eta(x,z;t)\rangle$ in a frame moving downstream at the {space- and time-averaged} surface velocity $U_{\rm int}$ (figure~\ref{fig:results}c). Here $r_x$ is an arbitrary streamwise increment, $\tau$ an arbitrary time increment, and angled brackets denote ensemble averaging in time, space, and across all equally-spaced samples. $\mathcal{C}$ is the physical-space counterpart of the spatiotemporal spectrum of the surface elevation.
The backward shift of $U_{\rm int}\tau$ applied to the streamwise increment compensates for Doppler shifting \citep{ferran-etal-2026}: perturbations passively advected by the mean velocity would then yield correlation peaks aligned along the vertical axis, whereas the peaks we observe depart from it, following the dispersion relation for gravity--capillary waves once written in terms of space and time lags and demonstrating that both gravitational and capillary dispersion mechanisms are active. Substituting $\kappa=2\pi/r_x$ into the dispersion relation~\eqref{eq:dispersion} and setting $\tau=2\pi/\omega_0$ gives the explicit correspondence represented by the red dash-dotted curve, $\tau(r_x) \;=\; 2\pi \sqrt{[\rho_2+\rho_1]/[(\rho_2-\rho_1)g2\pi/r_x+\sigma (2\pi / r_x)^3]}$.}

Despite the intrinsic anisotropy of the flow, surface modulation appears nearly isotropic even at the largest scales (figure~\ref{fig:results}d). We thus restrict our attention to the streamwise direction, and directly compare our analytical predictions with the surface spectra measured in the simulations. Note that, in principle, our predictions are formulated for isotropic surface spectra, yet streamwise spectra contain equal information and exhibit analogous scaling behaviour, as detailed in appendix~\S\ref{app:spectra}.
The power spectral density of the surface displacement, $S_\eta$ (figure~\ref{fig:results}e), displays two distinct scaling regimes separated by the critical wavenumber $\kappa_c$, given by equation~\eqref{eq:transition}. For $\kappa_x < \kappa_c$, $S_\eta \sim \kappa_x^{-7/3}$, consistent with the predicted gravitational limit and recent experimental observations  \citep{berhanu-falcon-2026}; for $\kappa_x > \kappa_c$, $S_\eta \sim \kappa_x^{-19/3}$, in agreement with the capillary limit (see table~\ref{tab:summary-scaling}). Finally, the spectra of the surface vertical velocity, $S_{\dot{\eta}}(\kappa_x)$ (figure~\ref{fig:results}f), also match our theoretical expectations in both regimes, with $S_{\dot{\eta}} \sim \kappa_x^{-1}$ and $S_{\dot{\eta}} \sim \kappa_x^{-5}$, respectively. These results confirm the validity of our theoretical framework within the regime where surface dynamics are faster than flow turbulence.

\subsection{Hyperelastic wall simulations: methods}

Approaching the regime where the surface is \textit{enslaved} by flow turbulence (\textcolor{black}{$\delta\ll1$}) is challenging in {fluid--fluid} systems, as reductions in density contrast or surface tension invariably cause the surface to break. To circumvent this limitation \textcolor{black}{and access the turbulence-enslaved regime, we therefore} shift our attention to the surface of a hyperelastic solid exposed to turbulent flow, shown in figure~\ref{fig:hyperelastic}. The constitutive model of the solid well describes the behaviour of a rubber coating or a gel layer, and good agreement is documented in literature with experimental setups~\citep{rosti-brandt-2017}.

Hyperelastic-wall simulations use the same numerical framework above, adapted to a fluid--solid configuration as introduced in former work \citep{koseki-aswathy-rosti-2025}. The data analysed here correspond to the elastic case with largest deformation in that study. The domain size is $6H \times (0.5+2)H \times 3H$, where the lower $0.5H$ is an incompressible neo-Hookean solid and the upper $2H$ is fluid, as in figure~\ref{fig:hyperelastic}. No-slip or penetration is imposed at the top boundary, while the streamwise and spanwise directions are periodic. A body force dynamically adjusted to maintain \textcolor{black}{a constant flow rate yielding} $Re = \rho_f U H/\mu_f \approx 2{,}800$ drives the turbulent flow~\citep[in the same fashion described by][]{hasegawa-quadrio-frohnapfel-2014}. Density and viscosity \textcolor{black}{are matched between the fluid and the solid}, and the solid shear modulus is $G/(\rho_f U^2) = 0.5$. Gravity and surface tension are absent.
 
The velocity is continuous across the fluid--solid interface, while the stress tensor in the solid follows the incompressible Mooney--Rivlin formulation \citep{bonet-wood-2008},
\begin{equation}
   \mathbf{\mathcal{T}}_s
   = \mu_s(\nabla\mathbf{u} + \nabla\mathbf{u}^T)
     + G\mathbf{\mathcal{B}},
\end{equation}
where $\mathbf{\mathcal{B}}$ is the left Cauchy--Green deformation tensor, evolved through
\begin{equation}
   \frac{\partial \mathbf{\mathcal{B}}}{\partial t}
   + \nabla\cdot(\mathbf{u}\mathbf{\mathcal{B}})
   = \mathbf{\mathcal{B}}\cdot\nabla\mathbf{u}
   + (\nabla\mathbf{u})^T\cdot\mathbf{\mathcal{B}}.
\end{equation}
Fluid and solid contributions are blended in each cell via the VOF method.

The equations are solved on a uniform grid of $1296 \times 540 \times 648$ points. After a transient of $\sim 300\,H/U$, statistics are collected over $\sim 80\,H/U$ at intervals of $1\,H/U$.

\subsection{Hyperelastic wall simulations: results}

\begin{figure}
   \centering
  \includegraphics[width=\textwidth]{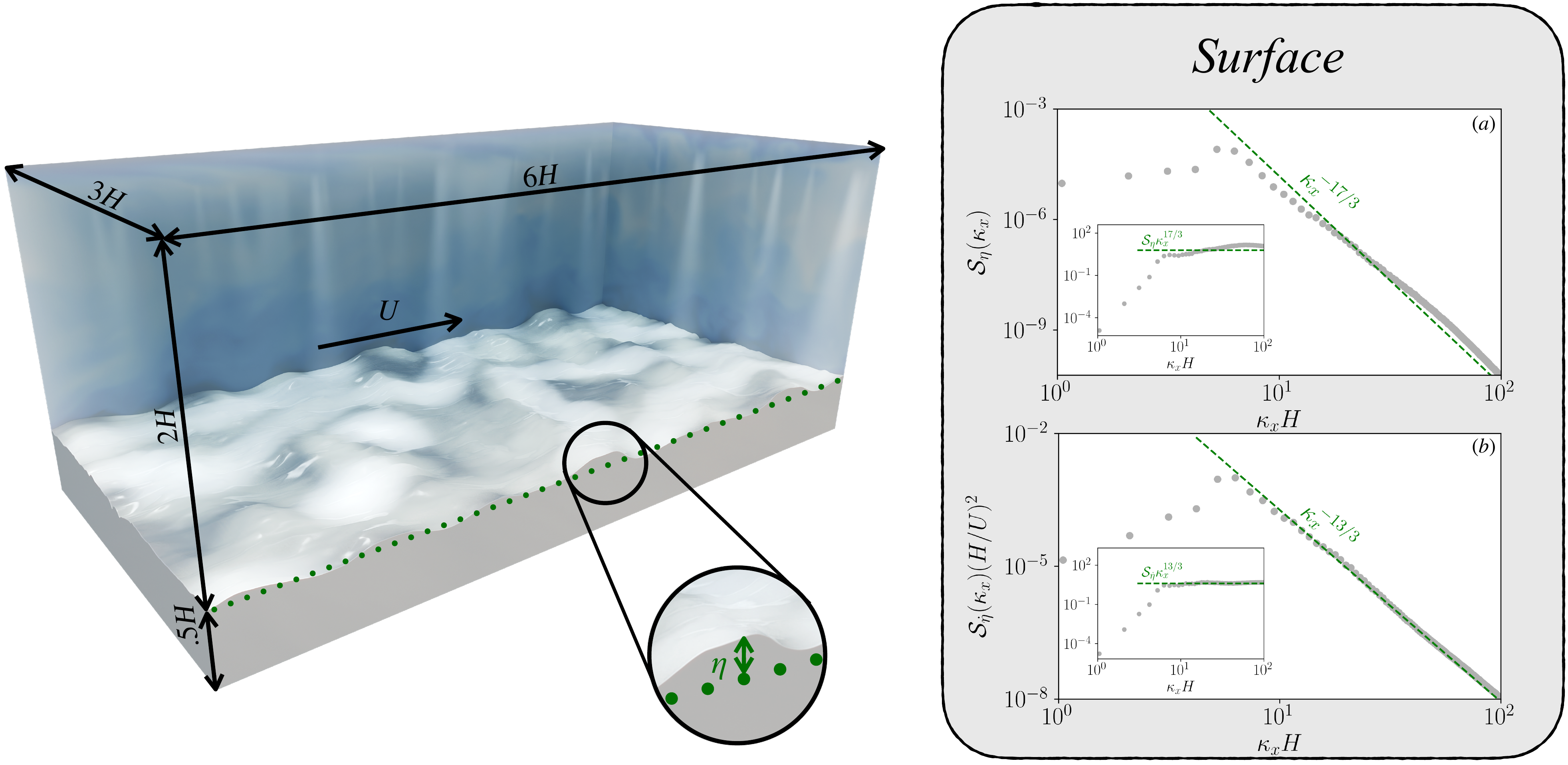}
  \caption{\textbf{Elastic solid surface in turbulence.} We consider a thick layer of rubber coating the bottom wall of a turbulent channel flow, as investigated in a previous work from our group \citep{koseki-aswathy-rosti-2025}. We thus compute the elevation (panel $a$) and vertical velocity spectra (panel $b$) of the solid surface along the streamwise direction. Remarkable agreement is found with our theoretical predictions in the regime where surface motion is enslaved by flow turbulence. Spectra compensated with their predicted scalings are reported in the insets.}
  \label{fig:hyperelastic}
\end{figure}

In the solid case, surface continuity is enforced by construction, whereas the absence of gravity and capillarity suppresses the fluid-interface restoring terms introduced in equation~\eqref{eq:dispersion}. This configuration is therefore not used to probe the detailed form of the restoring dynamics, but specifically to test the slow, viscous-dominated enslaved limit, for which the leading-order scaling is independent of the precise restoring mechanism. Moreover, because $\zeta \gtrsim 1$, viscous effects dominate the response on the turbulent timescale and the solid surface exhibits a viscous-dominated enslaved response whose spectral signatures exhibit remarkable agreement with our theoretical predictions: $S_{\eta} \sim \kappa_x^{-17/3}$ (figure~\ref{fig:hyperelastic}a) and $S_{\dot{\eta}} \sim \kappa_x^{-13/3}$ (figure~\ref{fig:hyperelastic}b). These observations confirm that our theoretical framework also holds in the complementary regime, where surface dynamics are slower than flow turbulence.


\section{Conclusions}
After introducing a comprehensive theoretical framework for free-surface dynamics in turbulence and validating it against state-of-the-art numerical simulations, several remarks are in order.

\textcolor{black}{Importantly, although our theory relies on a linearised representation of surface dynamics (equation~\ref{eq:osc}), the numerical simulations used for validation are not linearised. They solve the fully nonlinear and coupled two-phase Navier--Stokes equations in the presence of non-breaking waves of large amplitude, {comparable to those reported in wave turbulence studies} \citep{falcon-mordant-2022}, retaining the complete nonlinear interface dynamics. Nonlinear wave--wave interactions \citep{zakharov-filonenko-1967, newell-rumpf-2011, nazarenko-lukaschuk-2016,galtier-2021} are therefore not removed from the simulations. The agreement with the linear predictions is not imposed by the numerical model; it indicates a posteriori that nonlinear wave interactions do not provide the leading contribution to the frequency-integrated spectra in the configurations and over the time intervals considered. This does not imply that such interactions are strictly absent, but just sub-dominant in the regimes considered.
Our conclusions apply to the configurations studied here, in which the flow is strongly turbulent and continuously feeds the surface with broadband stochastic forcing. We do not claim that wave turbulence cannot develop at a deformable interface in general: e.g., weak-turbulence theory is known to apply when the fluids on both sides are quiescent or only weakly disturbed, so that the interface behaves as an almost autonomous weakly nonlinear wave system \citep{deike-etal-2014,bonnefoy-etal-2016}.}

\begin{figure}
  \centering
  \includegraphics[width=\textwidth]{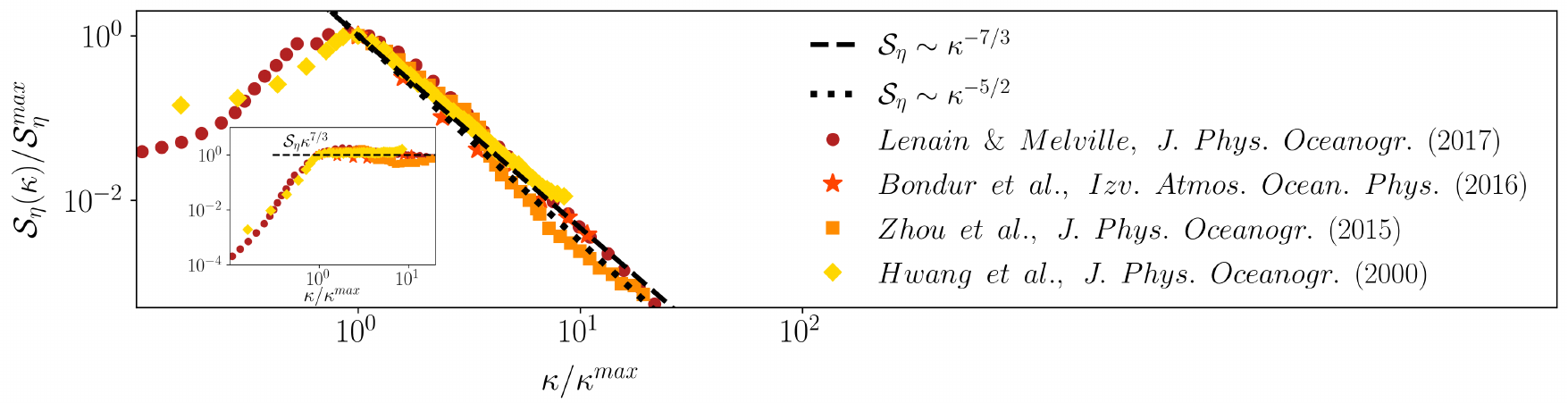}
  \caption{\textbf{Displacement spectrum from aerial measurements of the oceanic surface compared to analytical predictions.} Measurements from literature \citep{hwang-etal-2000, zhou-etal-2015, Bondur-etal-2016, lenain-melville-2017} are shown in colour, while the black dashed line denotes our theoretical prediction, and the black dotted line the prediction from wave turbulence. All spectra (including those compensated with our predicted scaling in the inset) are normalised by their maximum value, $S_\eta^{\max}$, and wavenumbers by $\kappa^{\max}$, the wavenumber at which $S_\eta^{\max}$ occurs.}
  \label{fig:oceanic}
\end{figure}

When intrinsic surface dynamics emerge, some of our spectral exponents are numerically close to the classical gravity-range spectra used in oceanography \citep{toba-1973-2, philips-1985}. In particular, our prediction $S_\eta(\kappa)\sim\kappa^{-7/3}$ in the gravity-dominated regime is difficult to distinguish, over a limited inertial range, from the weak-turbulence result $S_\eta(\kappa)\sim\kappa^{-5/2}$ for gravity waves \citep{zakharov-filonenko-1967}. Aerial measurements of the oceanic surface \citep[][in figure~\ref{fig:oceanic}]{hwang-etal-2000, zhou-etal-2015, Bondur-etal-2016, lenain-melville-2017} can be interpreted as compatible with both the predictions of our theory and of wave turbulence. 

In the capillary regime, however, the situation is different: our prediction $S_\eta(\kappa)\sim\kappa^{-19/3}$ is much steeper than the weak-turbulence scaling $S_\eta(\kappa)\sim\kappa^{-15/4}$ for the same one-dimensional isotropic spectrum \citep{zakharov-filonenko-1967}. The mismatch with wave turbulence theory is even more pronounced for the vertical-velocity spectrum. In the capillary regime, weak wave turbulence predicts $S_{\dot\eta}(\kappa)\propto\kappa^{-3/4}$ (and an even shallower slope for gravity waves) \citep{zakharov-filonenko-1967}, whereas our theory and simulations yield $S_{\dot\eta}(\kappa)\sim\kappa^{-1}$ and $S_{\dot\eta}(\kappa)\sim\kappa^{-5}$ in the gravity- and capillary-dominated ranges, respectively (table~\ref{tab:summary-scaling}). As a result, matching the slope of $S_\eta(\kappa)$ alone, especially in the gravity range, where several exponents are numerically close and have already been reported in former investigations \citep{ryabkova-etal-2019,giamagas-etal-2023}, is not sufficient to determine the dominant surface processes. {Instead, the combined evidence from the pressure spectrum, $S_p$ (or $S_{\Delta p}$ in the theory), together with $S_\eta$ and $S_{\dot\eta}$, might provide a clearer hint towards dynamics controlled by the turbulent pressure forcing or by a weak-turbulence cascade.}

 {An additional feature stemming from our theory is that, when surface dynamics are enslaved by flow turbulence in the inertia-dominated limit, the interface simply accommodates the pressure and velocity fluctuations imposed by the bulk, without appreciably feeding back on the flow.} In this regime, the vertical-velocity spectrum of the surface, \( S_{\dot\eta}(\kappa) \sim \kappa^{-5/3}, \) reproduces the Kolmogorov velocity spectrum of the turbulent bulk across the inertial interval (both above and below the capillary scale). Detecting the $\kappa^{-5/3}$ scaling in $S_{\dot\eta}(\kappa)$ thus confirms that the surface behaves as a passive free boundary driven by bulk turbulence, enabling direct access to inertial-range velocity statistics without intrusive volumetric measurements.

Introducing the median shear contribution to surface dynamics compared to pressure, here defined as the ratio of their respective root-mean-squared values $R=\tau_{\rm rms}/P_{\rm rms}$, we confirm that $R\sim\mathcal{O}(10^{-2})$ in both our simulated scenarios. Our predictions might therefore come short in shear-dominated conditions, like those arising when atmospheric winds generate oceanic waves \citep{zdyrski-feddersen-2021, wu-popinet-deike-2022, matsuda-etal-2023} or when they blow over vegetation canopies \citep{foggirota-etal-2024-2, lohrer-frohlich-2025}, making them sway. Further understanding is necessary to fully describe those circumstances.

In summary, we have developed a general theoretical framework for free-surface dynamics in turbulence (i.e., turbulence-induced-waves) and identified two distinct regimes: one in which intrinsic surface dynamics dominate (e.g., capillary or gravity waves) and one in which the surface is enslaved by the turbulent flow. 
Our theory, validated against our state-of-the-art numerical simulations, assumes that surface motion is driven primarily by pressure fluctuations and thus requires extension to describe shear-dominated conditions (e.g. wind-driven waves). \textcolor{black}{It might nevertheless benefit the characterisation of geophysically relevant scenarios dominated by flow- (instead of wave-) turbulence, such as interfacial motions over an intense submerged current or upwelling, the wake of a ship, or the surface of a rapidly flowing stream.}
It also remains of fundamental interest to clarify the respective conditions under which turbulence-induced-waves as in our study, wave-induced-turbulence \citep{babanin-2006}, and wave turbulence due to resonant wave--wave interactions \citep{newell-rumpf-2011} become significant.

\backsection[Acknowledgements]{This research was supported by the Okinawa Institute of Science and Technology Graduate University (OIST) with subsidy funding to M.E.R. from the Cabinet Office, Government of Japan. M.E.R.~also acknowledges funding from the Japan Society for the Promotion of Science (JSPS), grants 24K00810 and 24K17210. G.F.R. and M.E.R.~acknowledge the computer time provided by the Scientific Computing and Data Analysis section of the Core Facilities at OIST and the computational resources offered by the HPCI System Research Project with grants hp220402 and hp250035. The authors acknowledge Morie Koseki for providing and discussing the elastic solid data, published in \citet{koseki-aswathy-rosti-2025}.}

\backsection[Declaration of interests]{The authors declare there are no conflicts of interest for this work.}

\backsection[Data availability statement]{All data needed to evaluate the conclusions of this work are present in the main text and/or the appendices. Full details of the code implementation, validation, and related resources are available on the website of the Complex Fluids and Flows Unit at OIST (\url{www.oist.jp/research/research-units/cffu/fujin}).}

\backsection[Author ORCIDs]{\\ \\
\orcidA{} Giulio Foggi Rota \url{https://orcid.org/0000-0002-4361-6521}; \\
\orcidB{} Andrea Mazzino \url{https://orcid.org/0000-0003-0170-2891}; \\
\orcidC{} Marco Edoardo Rosti \url{https://orcid.org/0000-0002-9004-2292}. \\
}


\appendix

\begingroup\color{black}
\section{Derivation of the linearised interface dynamics}
\label{app:osc}

Equation~\eqref{eq:osc} is the standard leading-order reduction of the
two-phase free-boundary problem obtained by linearising about a flat
interface, under the deep-water and small-slope assumptions.
 
Importantly, this reduction does not assume that the turbulent flows on
the two sides of the interface are irrotational. The fluctuating pressure
jump $\Delta p'$ generated by those flows is retained as an external
forcing. Only the small interface-induced response is described through
an irrotational perturbation,
\begin{equation}
\boldsymbol{u}^{\,w}_i=\boldsymbol{\nabla}\phi_i,
\qquad
\nabla^2\phi_i=0,
\end{equation}
where $i=1,2$ labels the upper and lower phases, respectively.
 
Let the unperturbed interface be at $y=0$, with fluid~1 above and
fluid~2 below, and consider a horizontal Fourier mode of modulus
$\kappa$. In a frame moving with the mean interfacial velocity, the
linearised kinematic condition and decay away from the interface give
\begin{equation}
\widehat{\phi}_{1}
=
-\frac{\dot{\widehat{\eta}}_{\boldsymbol{\kappa}}}{\kappa}
e^{-\kappa y},
\qquad y>0,
\end{equation}
and
\begin{equation}
\widehat{\phi}_{2}
=
\frac{\dot{\widehat{\eta}}_{\boldsymbol{\kappa}}}{\kappa}
e^{\kappa y},
\qquad y<0.
\end{equation}
 
Let $\widehat{p}^{\,(1)}_{i,\boldsymbol{\kappa}}$ denote the pressure
perturbation at the displaced interface relative to the flat
equilibrium. Within the adopted reduction, the linearised Bernoulli
pressure and the hydrostatic correction give
\begin{equation}
\widehat{p}^{\,(1)}_{i,\boldsymbol{\kappa}}
=
\widehat{p}^{\, \prime}_{i,\boldsymbol{\kappa}}
-\rho_i\partial_t\widehat{\phi}_i(0,t)
-\rho_i g\widehat{\eta}_{\boldsymbol{\kappa}},
\end{equation}
where $\widehat{p}^{\, \prime}_{i,\boldsymbol{\kappa}}$ is the prescribed turbulent
pressure fluctuation evaluated at $y=0$. Defining the forcing jump as
$\widehat{\Delta p'}_{\boldsymbol{\kappa}}
=\widehat{p}^{\, \prime}_{1,\boldsymbol{\kappa}}
-\widehat{p}^{\, \prime}_{2,\boldsymbol{\kappa}}$ and substituting the two
potentials yields
\begin{equation}
\widehat{p}^{\,(1)}_{1,\boldsymbol{\kappa}}
-
\widehat{p}^{\,(1)}_{2,\boldsymbol{\kappa}}
=
\widehat{\Delta p'}_{\boldsymbol{\kappa}}
+
\frac{\rho_1+\rho_2}{\kappa}
\ddot{\widehat{\eta}}_{\boldsymbol{\kappa}}
+
(\rho_2-\rho_1)g\widehat{\eta}_{\boldsymbol{\kappa}}.
\end{equation}
 
The normal viscous stresses follow from
$u^{\,w}_{y,i}=\partial_y\phi_i$. In particular,
\begin{equation}
\left.
\partial_y\widehat{u}^{\,w}_{y,1}
\right|_{y=0}
=
-\kappa\dot{\widehat{\eta}}_{\boldsymbol{\kappa}},
\qquad
\left.
\partial_y\widehat{u}^{\,w}_{y,2}
\right|_{y=0}
=
\kappa\dot{\widehat{\eta}}_{\boldsymbol{\kappa}}.
\end{equation}
With the unit normal directed from fluid~2 to fluid~1, the linearised
curvature is $\mathcal{K}\simeq-\nabla_\perp^2\eta$, where
$\nabla_\perp^2=\partial_x^2+\partial_z^2$, so that
$\widehat{\mathcal{K}}_{\boldsymbol{\kappa}}
=\kappa^2\widehat{\eta}_{\boldsymbol{\kappa}}$.
Retaining the normal viscous stresses of the potential response,
the linearised normal-stress condition therefore reads
\begin{equation}
\widehat{p}^{\,(1)}_{1,\boldsymbol{\kappa}}
-
\widehat{p}^{\,(1)}_{2,\boldsymbol{\kappa}}
-
2\mu_1
\left.
\partial_y\widehat{u}^{\,w}_{y,1}
\right|_{y=0}
+
2\mu_2
\left.
\partial_y\widehat{u}^{\,w}_{y,2}
\right|_{y=0}
+
\sigma\kappa^2\widehat{\eta}_{\boldsymbol{\kappa}}
=0,
\end{equation}
where $\mu_i=\rho_i\nu_i$ is the dynamic viscosity of phase~$i$.
Substitution of the expressions above yields
\begin{equation}
\frac{\rho_1+\rho_2}{\kappa}
\ddot{\widehat{\eta}}_{\boldsymbol{\kappa}}
+
2(\rho_1\nu_1+\rho_2\nu_2)\kappa
\dot{\widehat{\eta}}_{\boldsymbol{\kappa}}
+
\left[
(\rho_2-\rho_1)g+\sigma\kappa^2
\right]
\widehat{\eta}_{\boldsymbol{\kappa}}
=
-\widehat{\Delta p'}_{\boldsymbol{\kappa}},
\end{equation}
where the sign on the right-hand side follows the pressure-jump
convention adopted throughout this work. Multiplication by
$\kappa/(\rho_1+\rho_2)$ gives equation~\eqref{eq:osc}.
 
Equation~\eqref{eq:osc} is therefore not an ad hoc oscillator model: its inertial,
gravity, capillary, and pressure-forcing terms follow directly from the
linearised two-phase interfacial problem. The viscous term is the
corresponding viscous-potential-flow approximation, obtained by retaining
the leading normal viscous stress. The resulting equation is exact within
the inviscid linear potential-flow problem and approximate with respect
to viscosity. It does not include finite-depth corrections, modifications
due to mean shear, rotational viscous modes, or nonlinear wave--wave
interactions.
\endgroup

\section{Regime of motion in our quasi air-water simulations}
\label{app:specs}

\begin{figure} [h]
   \centering
   \includegraphics[width=.4\textwidth]{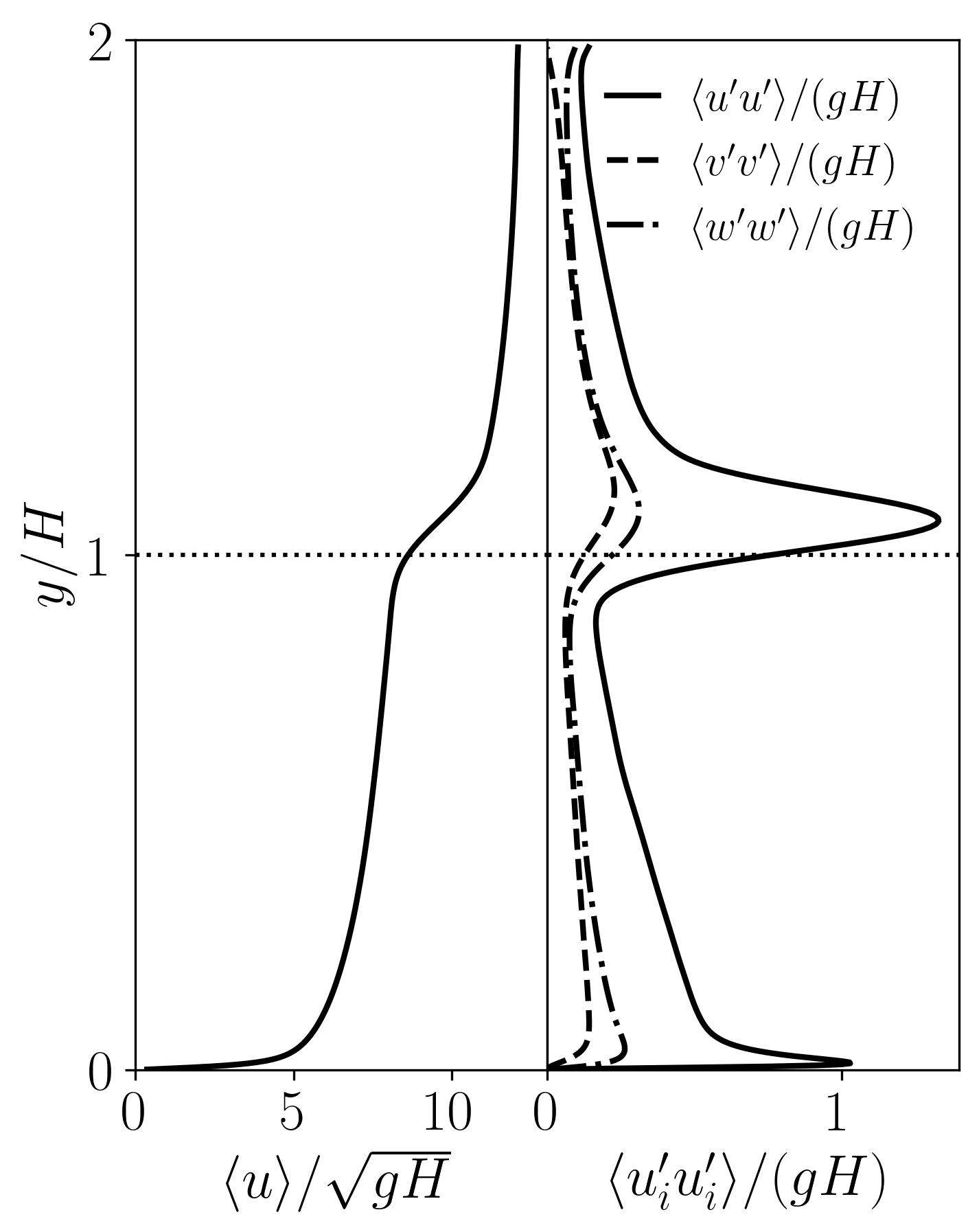}
   \caption{\textbf{Mean velocity (left) and fluctuation (right) profiles from our quasi-air/water simulation}, with the average interface position denoted by a dotted line.}
   \label{fig:profiles}
\end{figure}

\textcolor{black}{In our quasi air--water simulations, performed under the same forcing conditions assessed by \citet{foggirota-chiarini-rosti-2026} but in a larger computational domain, the lower fluid layer flows in an almost ``open channel'' configuration, while the upper fluid layer participates in the motion (i.e., the wind gently blows on the waves along the streamwise direction).
The mean profile of the streamwise velocity demonstrating this description is reported in the left panel of figure~\ref{fig:profiles}, while in the right panel we characterise the fluctuation intensity of the various velocity components. In the interface region, where fluctuations reach their maxima, we measure a turbulent dissipation rate of $\varepsilon_{\rm int} \sim 0.05\sqrt{g^3 H}$. The large-scale Froude number based on the turbulence intensity ($u_{\rm rms}=\sqrt{u_i^\prime u_i^\prime}$) is $Fr_H=\sqrt{{(\rho_2+\rho_1)u_{\rm rms}^2}/{[(\rho_2-\rho_1)gH]}}\approx1$, while $Fr_{\rm int}=\sqrt{{(\rho_2+\rho_1)U_{\rm int}^2}/{[(\rho_2-\rho_1)gH]}}\approx7$ is found based on the interface velocity. As $(\rho_2+\rho_1)/(\rho_2-\rho_1)\approx1$ given the high density ratio considered, these two definitions of the Froude number effectively coincide with the scaled values of the fluctuation and mean velocity profiles at the interface.}

\begin{figure} 
  \centering
  \includegraphics[width=.48\textwidth]{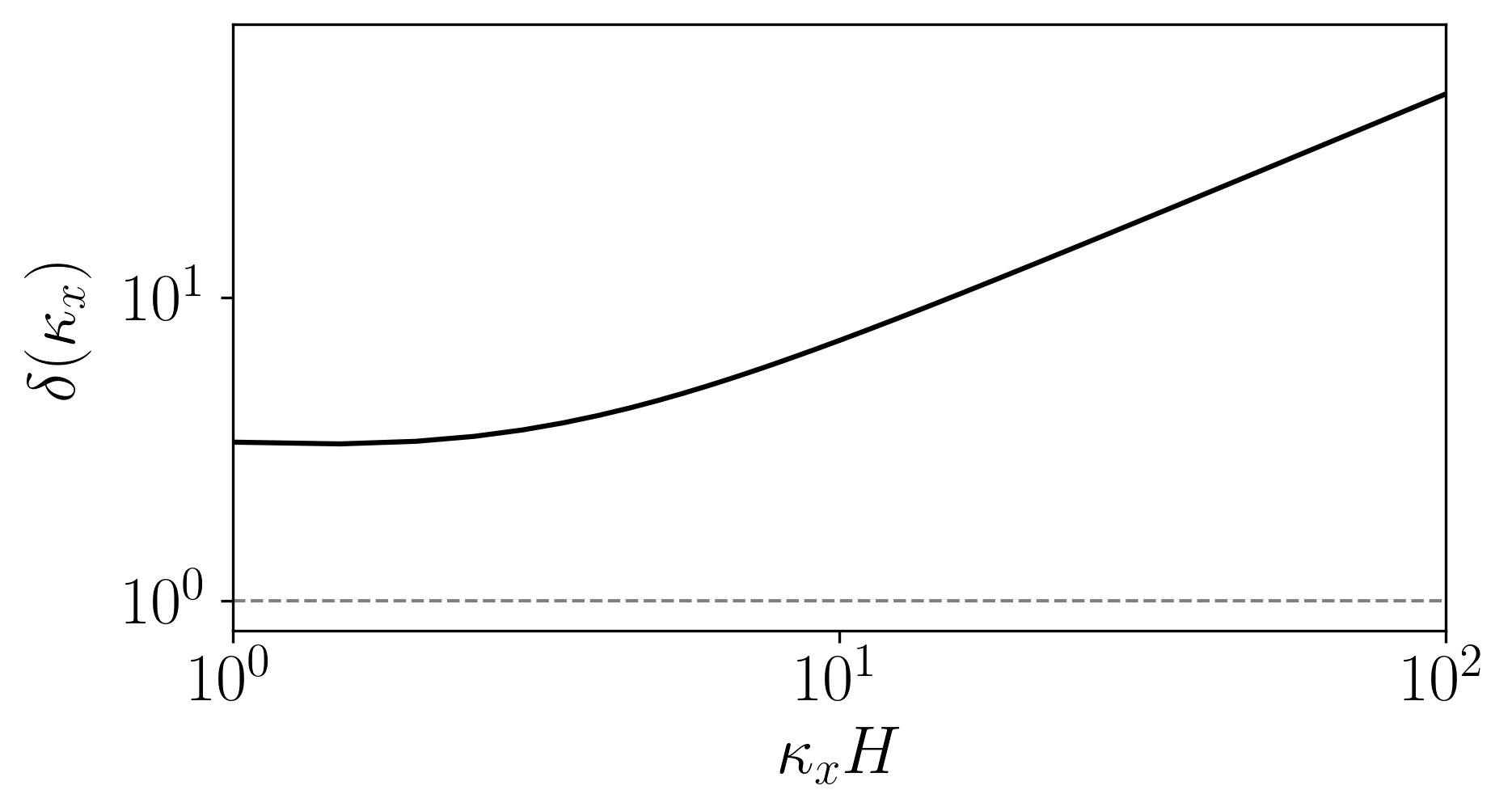}
  \includegraphics[width=.48\textwidth]{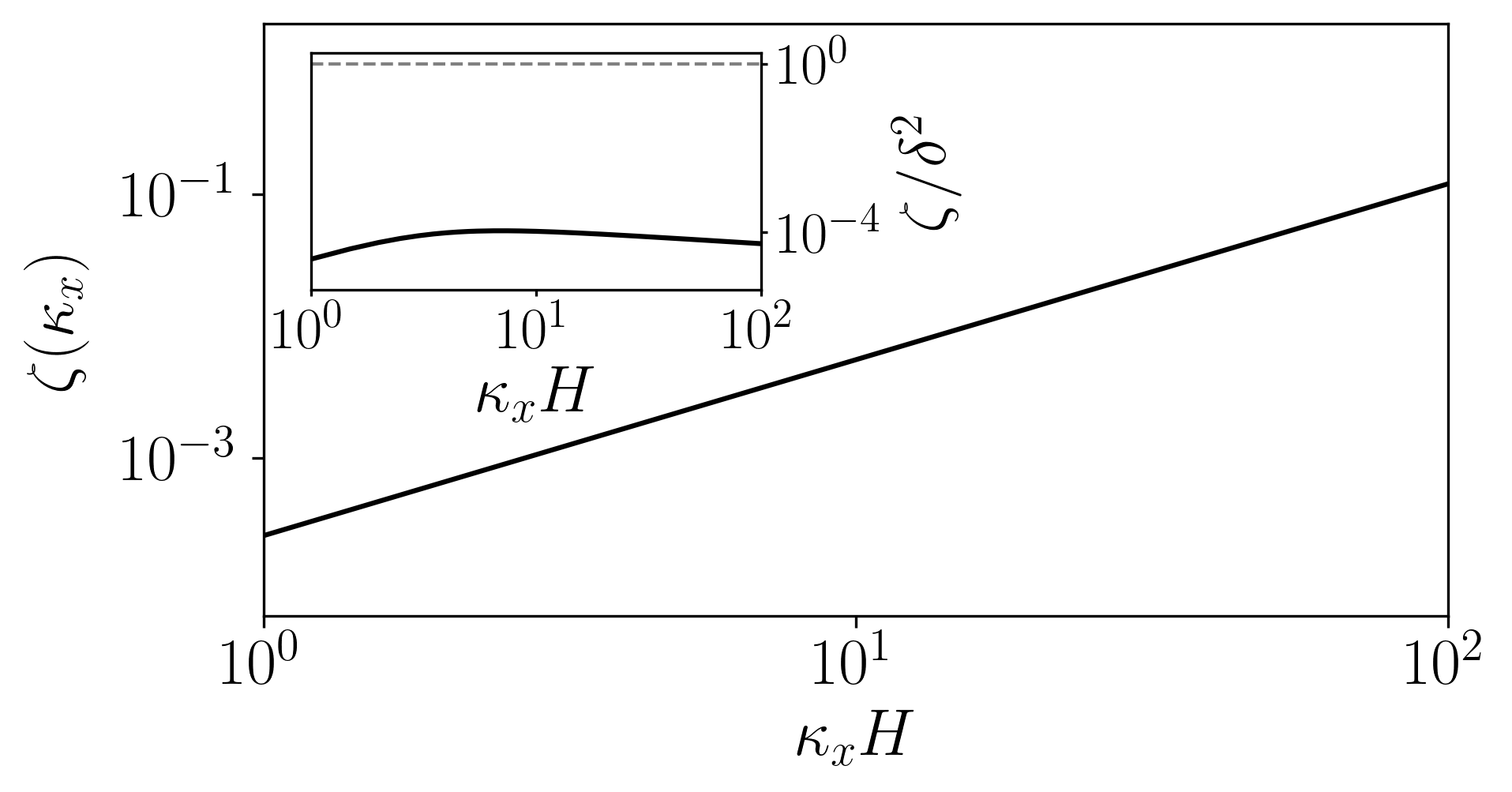}
  \caption{\textbf{Frequency ratios between \textcolor{black}{turbulent flow dynamics and intrinsic surface dynamics (left)/surface damping (right).}} \(\delta={\omega_0}/{\omega_t} >1\) \textcolor{black}{and \( \zeta = \gamma / \omega_t \ll \delta^2 \)} at all streamwise wavenumbers \( \kappa_x \), confirming the placement of the reported quasi air--water simulation in the regime where intrinsic surface dynamics are faster than the surrounding flow turbulence, and are thus able to emerge.}
  \label{fig:delta}
\end{figure}

To confirm that our quasi air--water simulations are performed in the flow regime where intrinsic surface dynamics emerge from the underlying flow turbulence, we compute the value of \(\delta={\omega_0}/{\omega_t}\) \textcolor{black}{and \(\zeta={\gamma}/{\omega_t}\)} as functions of the streamwise wavenumber $\kappa_x$, with \(\omega_0\), \(\omega_t\), \textcolor{black}{and \( \gamma \)} as defined in the main text.We indeed find \( \delta > 1\) and \textcolor{black}{\( \zeta \ll \delta^2 \)} throughout all the wavenumbers considered, as demonstrated in figure~\ref{fig:delta}, confirming the placement of the simulation reported in the main text within the intended regime of the theoretical framework.

\textcolor{black}{Experimentally, evaluating $\delta$ requires an estimate of the turbulent dissipation rate close to the interface, which is inherently more complex than determining bulk Froude or Weber numbers. Dissipation can be estimated from probe measurements of the velocity-gradients.}

\section{Equivalence of isotropic and one-dimensional elevation spectra}
\label{app:spectra}

We show here that, for a statistically homogeneous and horizontally
isotropic interface displacement $\eta(x,z,t)$, the azimuthally
integrated (isotropic) elevation spectrum
\begin{equation}
S_{\eta}(\kappa)=\int_{-\infty}^{\infty} S_{\eta}(\kappa,\omega)\,d\omega,
\label{eq:Siso_def}
\end{equation}
has the same scaling in $\kappa$ as the one-dimensional spectrum
obtained by Fourier transforming only along the $x$--direction and
averaging along $z$.

\subsection{Definitions in wavevector--frequency space}

Consistently with the main text, we decompose $\eta$ in horizontal
Fourier modes,
\begin{equation}
\eta(x,z,t)=\iint \hat{\eta}_{\bk}(t)\,
\mathrm{e}^{\mathrm{i}(\kappa_x x+\kappa_z z)}\,d^2\bk,
\qquad
\bk=(\kappa_x,\kappa_z),\quad
\kappa = |\bk|.
\label{eq:spatialFT}
\end{equation}
The temporal Fourier transform of each mode is $ \hat{\eta}_{\bk}(\omega)=\int_{-\infty}^{\infty} \hat{\eta}_{\bk}(t)\,\mathrm{e}^{\mathrm{i}\omega t}\,dt$.
For a statistically stationary and homogeneous field we define the
(directionally resolved) elevation power spectral density $F_{\eta}$
via
\begin{equation}
\left\langle
\hat{\eta}_{\bk}(\omega)\,\hat{\eta}^{*}_{\bk'}(\omega')
\right\rangle
=(2\pi)^3\delta(\bk-\bk')\,\delta(\omega-\omega')\,
F_{\eta}(\bk,\omega).
\label{eq:dirspec_def}
\end{equation}

Horizontal isotropy implies that $F_{\eta}$ depends on $\bk$ only
through its modulus $\kappa$, i.e.
$F_{\eta}(\bk,\omega)=F_{\eta}(\kappa,\omega)$.
The azimuthally integrated (isotropic) elevation spectrum in
$(\kappa,\omega)$--space is then
\begin{equation}
S_{\eta}(\kappa,\omega)
= \int_{0}^{2\pi} F_{\eta}(\kappa,\omega)\,\kappa\,d\theta
= 2\pi\kappa\,F_{\eta}(\kappa,\omega),
\label{eq:Siso_komega}
\end{equation}
where $\theta$ is the polar angle in the $(\kappa_x,\kappa_z)$ plane.
Integrating over frequency gives~\eqref{eq:Siso_def}.

\subsection{One-dimensional spectrum from Fourier transform in $x$}

We now construct the spectrum that is obtained in practice by
Fourier transforming only along $x$ and averaging along $z$.
For each fixed $z$ we define the partial transform
\begin{equation}
\tilde{\eta}(\kappa_x,z,t)
= \frac{1}{2\pi}\int_{-\infty}^{\infty}
\eta(x,z,t)\,\mathrm{e}^{-\mathrm{i}\kappa_x x}\,dx .
\label{eq:FTx_def}
\end{equation}
Substituting \eqref{eq:spatialFT} into \eqref{eq:FTx_def} and using
the standard Fourier identity
$\int \mathrm{e}^{\mathrm{i}(\kappa_x'-\kappa_x)x}dx
=2\pi\,\delta(\kappa_x'-\kappa_x)$ yields
\begin{equation}
\tilde{\eta}(\kappa_x,z,t)
= \int_{-\infty}^{\infty}
\hat{\eta}_{\bk}(t)\,\mathrm{e}^{\mathrm{i}\kappa_z z}\,d\kappa_z .
\label{eq:tilde_eta_kx}
\end{equation}
The temporal Fourier transform of $\tilde{\eta}$ is
\begin{equation}
\tilde{\eta}(\kappa_x,z,\omega)
= \int_{-\infty}^{\infty}
\tilde{\eta}(\kappa_x,z,t)\,\mathrm{e}^{\mathrm{i}\omega t}\,dt
= \int_{-\infty}^{\infty}
\hat{\eta}_{\bk}(\omega)\,\mathrm{e}^{\mathrm{i}\kappa_z z}\,d\kappa_z .
\label{eq:tilde_eta_kxomega}
\end{equation}

\begingroup\color{black}
We now construct the one-dimensional spectrum associated with a Fourier
transform along the $x$ direction. For a statistically homogeneous field,
the spectrum is independent of the spanwise position $z$; in numerical
practice, averaging along $z$ is used only to improve statistical
convergence.
 
For any fixed $z$, we define the one-dimensional elevation spectrum as
the coefficient of the remaining Dirac distributions,
\begin{equation}
\begin{aligned}
\left\langle
\widetilde{\eta}(\kappa_x,z,\omega)
\widetilde{\eta}^{\,*}(\kappa_x',z,\omega')
\right\rangle
={}&
(2\pi)^3
\delta(\kappa_x-\kappa_x')
\delta(\omega-\omega')
\\
&\times
S_{\eta}^{1D}(\kappa_x,\omega).
\end{aligned}
\label{eq:covariance}
\end{equation}
 
Using equation~\eqref{eq:tilde_eta_kxomega} and keeping distinct wavenumbers and frequencies
until the spectral covariance~\eqref{eq:dirspec_def} is applied, we obtain
\begin{align}
&
\left\langle
\widetilde{\eta}(\kappa_x,z,\omega)
\widetilde{\eta}^{\,*}(\kappa_x',z,\omega')
\right\rangle
\nonumber\\
&\quad =
\int_{-\infty}^{\infty}
\int_{-\infty}^{\infty}
e^{i(\kappa_z-\kappa_z')z}
\left\langle
\widehat{\eta}_{(\kappa_x,\kappa_z)}(\omega)
\widehat{\eta}^{\,*}_{(\kappa_x',\kappa_z')}(\omega')
\right\rangle
\,d\kappa_z\,d\kappa_z'
\nonumber\\
&\quad =
(2\pi)^3
\delta(\kappa_x-\kappa_x')
\delta(\omega-\omega')
\int_{-\infty}^{\infty}
F_{\eta}(\kappa_x,\kappa_z,\omega)
\,d\kappa_z .
\label{eq:explCovariance}
\end{align}
Here, the factor $\delta(\kappa_z-\kappa_z')$ contained in
$\delta(\boldsymbol{\kappa}-\boldsymbol{\kappa}')$ performs the
$\kappa_z'$ integration.
 
Comparing equations~\eqref{eq:covariance} and~\eqref{eq:explCovariance} gives
\begin{equation}
S_{\eta}^{1D}(\kappa_x,\omega)
=
\int_{-\infty}^{\infty}
F_{\eta}(\kappa_x,\kappa_z,\omega)
\,d\kappa_z .
\label{eq:Spec1D_from_F}
\end{equation}
Thus, the one-dimensional spectrum is obtained by integrating the
directionally resolved spectrum over the transverse wavenumber.
\endgroup

\subsection{Relation to the isotropic spectrum}

Under horizontal isotropy,
$F_{\eta}(\kappa_x,\kappa_z,\omega)=F_{\eta}(\kappa,\omega)$ with
$\kappa=(\kappa_x^2+\kappa_z^2)^{1/2}$.
Introducing polar coordinates in wavevector space and using
\eqref{eq:Siso_komega}, the integral
\eqref{eq:Spec1D_from_F} can be written as
\begin{equation}
S_{\eta}^{1\mathrm{D}}(\kappa_x,\omega)
=
\int_{-\infty}^{\infty}
F_{\eta}\!\left(\sqrt{\kappa_x^2+\kappa_z^2},\omega\right)
d\kappa_y
=
\frac{1}{\pi}\int_{|\kappa_x|}^{\infty}
\frac{S_{\eta}(\kappa,\omega)}{\sqrt{\kappa^2-\kappa_x^2}}\,d\kappa .
\label{eq:Abel_komega}
\end{equation}
Equation~\eqref{eq:Abel_komega} is an Abel-type transform relating the
azimuthally integrated isotropic spectrum $S_{\eta}(\kappa,\omega)$ to
the one-dimensional spectrum $S_{\eta}^{1\mathrm{D}}(\kappa_x,\omega)$.

We are interested in spectra integrated over frequency.  Defining
\begin{equation}
S_{\eta}^{1\mathrm{D}}(\kappa_x)
= \int_{-\infty}^{\infty}
S_{\eta}^{1\mathrm{D}}(\kappa_x,\omega)\,d\omega ,
\label{eq:Spec1D_k_def}
\end{equation}
and using \eqref{eq:Abel_komega}, we obtain
\begin{equation}
S_{\eta}^{1\mathrm{D}}(\kappa_x)
=
\frac{1}{\pi}\int_{|\kappa_x|}^{\infty}
\frac{S_{\eta}(\kappa)}{\sqrt{\kappa^2-\kappa_x^2}}\,d\kappa ,
\label{eq:Abel_k}
\end{equation}
with $S_{\eta}(\kappa)$ given by \eqref{eq:Siso_def}.  Thus the
frequency-integrated one-dimensional spectrum and the frequency-integrated
isotropic spectrum are related by the same Abel transform.

\subsection{Equivalence of scaling laws}

Assume that, in a given inertial range enclosed by a minimum and a maximum wavenumber $\kappa_{\min}$ and $\kappa_{\max}$, the isotropic spectrum obeys
a power law
\begin{equation}
S_{\eta}(\kappa) \sim C\,\kappa^{-n},
\qquad \kappa_{\min}\ll \kappa \ll \kappa_{\max},
\label{eq:Siso_scaling}
\end{equation}
with a constant pre-factor $C$ and exponent $n>0$.
Substituting \eqref{eq:Siso_scaling} into \eqref{eq:Abel_k} gives
\begin{equation}
S_{\eta}^{1\mathrm{D}}(\kappa_x)
\sim
\frac{C}{\pi}
\int_{\kappa_x}^{\infty}
\frac{\kappa^{-n}}{\sqrt{\kappa^2-\kappa_x^2}}\,d\kappa .
\label{eq:Spec1D_scaling_step1}
\end{equation}
Rescaling the integration variable as $\kappa=\kappa_x u$ yields
\begin{equation}
S_{\eta}^{1\mathrm{D}}(\kappa_x)
\sim
\frac{C}{\pi}\,\kappa_x^{-n}
\int_{1}^{\infty}
\frac{u^{-n}}{\sqrt{u^2-1}}\,du .
\label{eq:Spec1D_scaling_step2}
\end{equation}
The integral in \eqref{eq:Spec1D_scaling_step2} is a finite constant
that depends only on $n$ (and on the large--$\kappa$ cut-off implied
by the inertial range), so that
\begin{equation}
S_{\eta}^{1\mathrm{D}}(\kappa_x)
\propto \kappa_x^{-n},
\qquad
\kappa_{\min}\ll \kappa_x \ll \kappa_{\max}.
\label{eq:Spec1D_scaling}
\end{equation}

Equations~\eqref{eq:Siso_scaling} and \eqref{eq:Spec1D_scaling}
demonstrate that, under horizontal isotropy, the azimuthally
integrated elevation spectrum $S_{\eta}(\kappa)$ and the one-dimensional
spectrum $S_{\eta}^{1\mathrm{D}}(\kappa_x)$ obtained by Fourier
transform in $x$ and averaging in $z$ share the same scaling exponent
in wavenumber.  Differences between the two spectra affect only the
pre-factor (a constant depending on $n$), not the $\kappa$-dependence.
In virtue of this last observation, in the main text, we drop the $1D$ apex and simply report $S_{\eta}(\kappa_x)$, exhibiting the same scaling behaviour of $S_{\eta}(\kappa)$ (same for $S_{\dot\eta}$).

\bibliographystyle{jfm}
\bibliography{Wallturb}

\end{document}